\documentclass{ar-astro2e}
\usepackage{amsmath ,amssymb}
\usepackage{natbib}
\usepackage{graphicx} 
\newcommand{\avg}[1]{\left< #1 \right>} 

\begin{document}
\input epsf.tex    

\input psfig.sty

\jname{Annu. Rev. Astron. Astrophys}
\jyear{2010}
\jvol{49}
\ARinfo{1056-8700/97/0610-00}

\title{The Astrophysics of Ultrahigh Energy Cosmic Rays}

\markboth{Kotera \& Olinto}{Astrophysics of UHECRs}

\author{Kumiko Kotera and Angela V. Olinto
\affiliation{Department of Astronomy and Astrophysics,\\
Kavli Institute for Cosmological Physics, \\The University of Chicago, Chicago, IL 60637, USA}}


\begin{abstract}
The origin of the highest energy cosmic rays is still unknown. The discovery of their sources will reveal the workings of the most energetic astrophysical accelerators in the universe. Current observations show a spectrum consistent with an origin in extragalactic astrophysical sources. Candidate sources range from the birth of compact objects to explosions related to gamma-ray bursts or to events in active galaxies. We discuss the main effects of propagation from cosmologically distant sources including interactions with cosmic background radiation and magnetic fields. We examine possible acceleration mechanisms leading to a survey of candidate sources and their signatures. New questions arise from an observed hint of sky anisotropies and an unexpected evolution of composition indicators. Future observations may reach the necessary sensitivity to achieve charged particle astronomy and to observe ultrahigh energy photons and neutrinos, which will further illuminate the workings of the universe at these extreme energies. In addition to fostering a new understanding of high-energy astrophysical phenomena, the study of ultrahigh energy cosmic rays can constrain the structure of the Galactic and extragalactic magnetic fields as well as probe particle interactions at energies orders of magnitude higher than achieved in terrestrial accelerators.

\end{abstract}

\maketitle

\section{Introduction}

The observation that cosmic rays can exceed $10^{20}$ eV poses some interesting and challenging questions: Where do they come from? How can they be accelerated to such high energies? What kind of particles are they? What is the spatial distribution of their sources? What do they tell us about these extreme cosmic accelerators? How strong are the magnetic fields that they traverse on their way to Earth? How do they interact with the cosmic background radiation? What secondary particles are produced from these interactions? What can we learn about particle interactions at these otherwise inaccessible energies? Here we review recent progress towards answering these questions. 
 
\begin{figure}[!t]
\centerline{\includegraphics[height=0.8\textwidth]{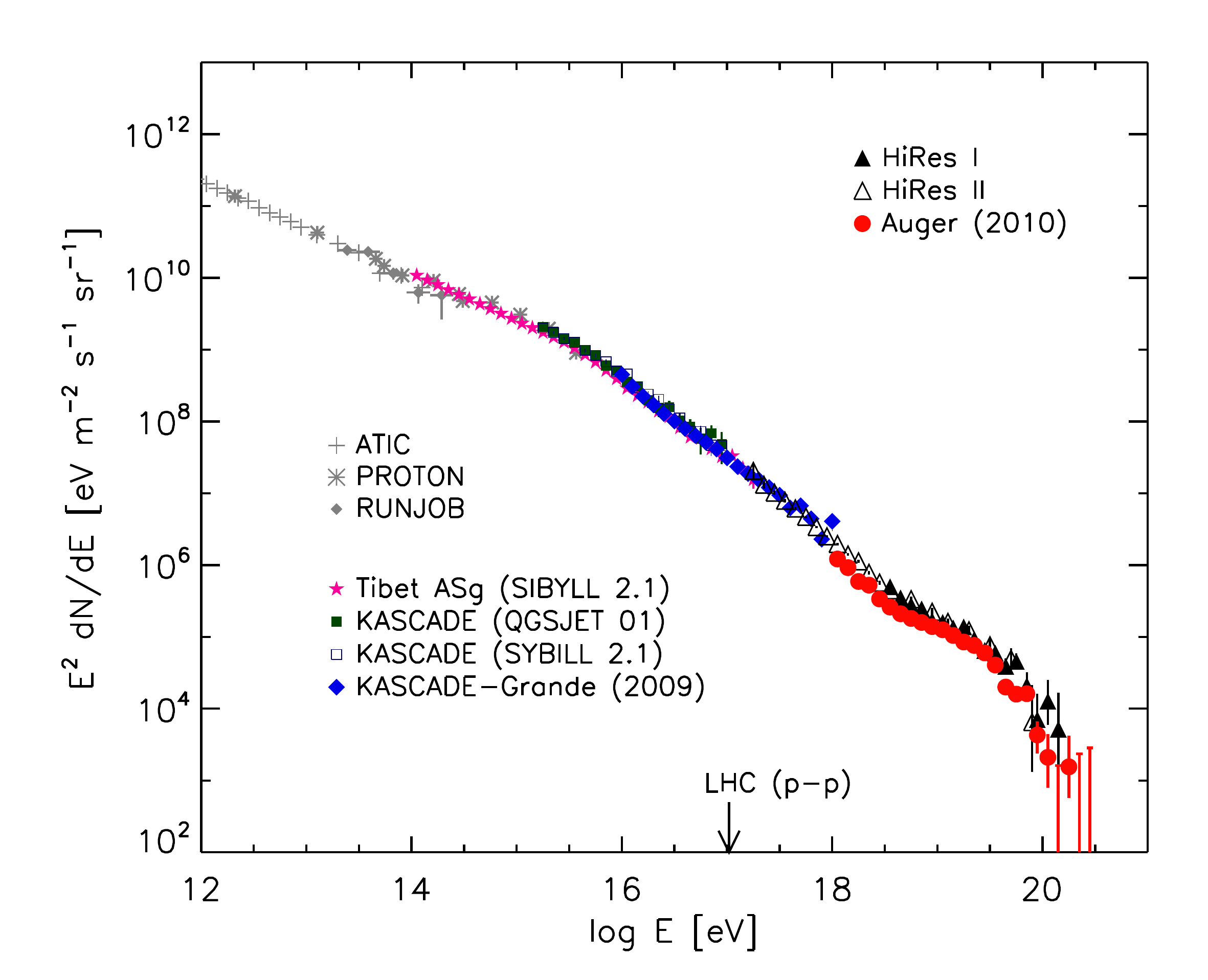}}
\caption{All particle cosmic ray flux multiplied by $E^2$ observed by ATIC \citep{Ahn08}, Proton \citep{Grigorov71}, RUNJOB \citep{Apanasenko01}, Tibet AS-$\gamma$ \citep{Chen08}, KASCADE \citep{Kampert04}, KASCADE-Grande \citep{KASCADE-Grande09}, HiRes-I \citep{Abbasi09}, HiRes-II \citep{Abbasi08}, and Auger \citep{Abraham10}.  LHC energy reach of $p-p$ collisions (in the frame of a proton) is indicated for comparison.}
\label{figure1}
\end{figure}

The dominant component of cosmic rays observed on Earth originate in the Galaxy. As shown in Figure \ref{figure1}, the study of this striking non-thermal spectrum requires a large number of instruments to cover over 8 orders of magnitude in energy and 24 in flux. Galactic cosmic rays are likely to originate in supernova remnants (see, e.g.,  \citealp{Hillas06}, for a recent update on the origin of Galactic cosmic rays). A transition from Galactic to extragalactic cosmic rays should occur somewhere between 1 PeV ($\equiv 10^{15}$ eV) and 1 EeV ($\equiv 10^{18}$ eV). Progress on determining this transition relies both on the study of the highest energies reached in Galactic accelerators as well as the search for extragalactic accelerators that produce ultrahigh energy cosmic rays (UHECRs). 

\begin{figure}[!t]
\centerline{\includegraphics[height=0.5\textheight]{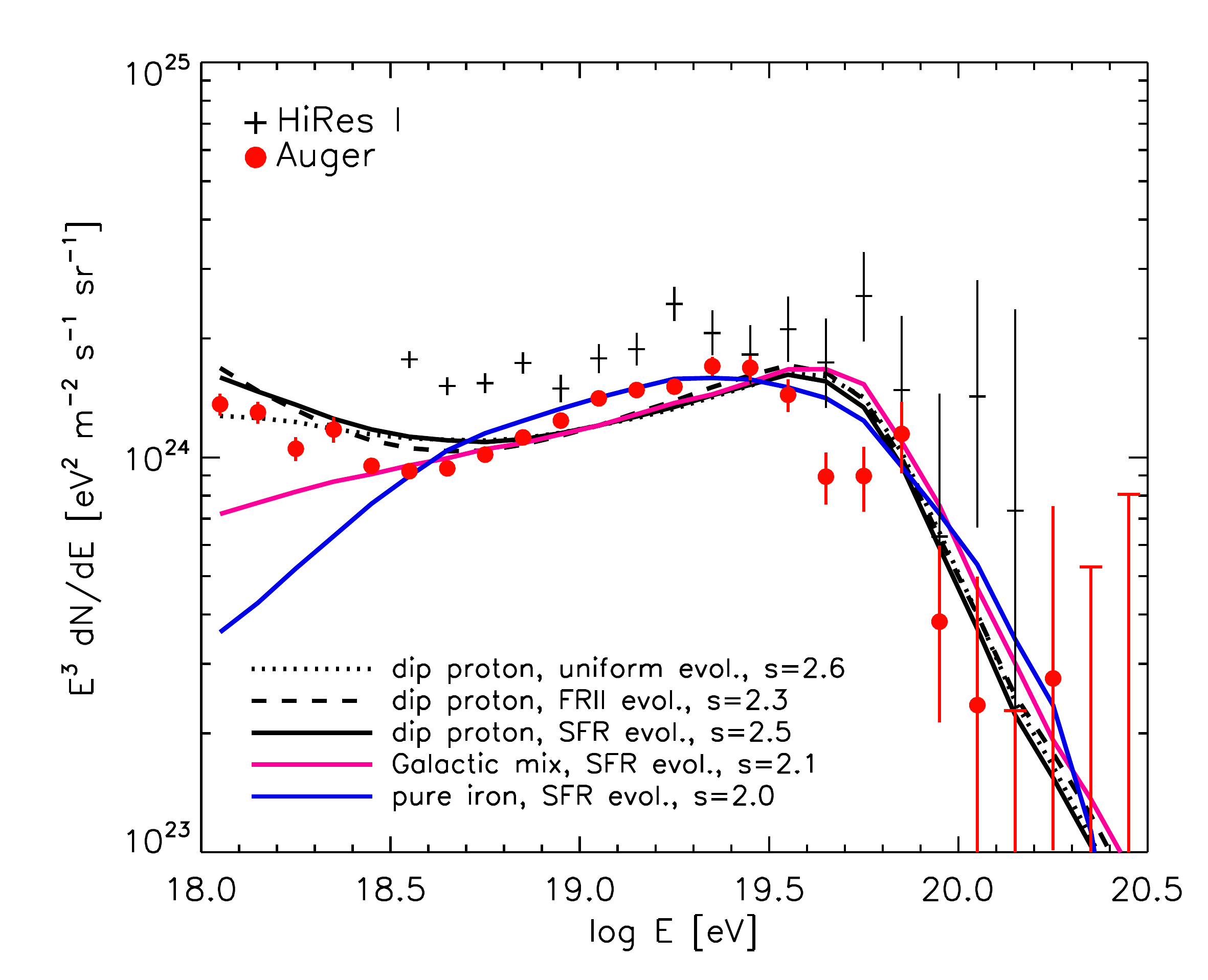}}
\caption{Spectrum of UHECRs multiplied by $E^3$ observed by HiRes I \citep{Abbasi09} and Auger \citep{Abraham10}. Overlaid are simulated spectra obtained for different models of the Galactic to extragalactic transition and different injected chemical compositions and spectral indices, $s$, described in Section~\ref{subsection:spectrum} and \ref{section:transition}. }
\label{figure2}
\end{figure}

We begin with a brief summary of recent observations (Section \ref{section:observation}), which reveal a spectrum whose shape supports the long-held notion that sources of UHECRs are extragalactic. As shown in Figure  \ref{figure2}, the crucial spectral feature recently established at the highest energies  is a steep decline in flux above about 30 EeV. This feature is reminiscent of the effect of interactions between extragalactic cosmic rays and the cosmic background radiation, named the Greisen-Zatsepin-Kuzmin (GZK) cutoff \citep{G66,ZK66}. Another important feature shown in Figure  \ref{figure2}  is the hardening of the spectrum at a few EeV, called the {\it ankle}, which may be caused by the transition from Galactic to extragalactic cosmic rays or by propagation losses if UHECRs are mostly protons.

As discussed in Section \ref{section:observation}, recent reports of a trend toward a heavier composition from a few EeV up to 40 EeV together with hints of anisotropies in the sky distribution above 60 EeV raise new and unexpected puzzles. An anisotropic sky distribution is expected for trans-GZK energies (i.e., energies above 60 EeV), if UHECRs are mainly protons, due to a combination of the GZK effect (which limits trans-GZK observed sources to lie within a few 100 Mpc), the anisotropic distribution of source bearing galaxies on 100 Mpc scales, and the low magnetic deflection of light trans-GZK nuclei by the Galactic and extragalactic magnetic fields. Therefore, the report of correlations between UHECRs above 55 EeV and the distribution of nearby active galaxies \citep{Auger1}  can be simply interpreted as protons from nearby sources within the so-called GZK sphere.  However, composition indicators from shower development observations argue for a transition to a heavier component from a few EeV up to 40 EeV \citep{Abraham:2010yv}. Heavy nuclei dominated injection models are quite rare in the astrophysical literature of candidate sources (see Section \ref{section:sources}) and if iron is the main component at the highest energies, Galactic magnetic fields should wash out most anisotropic patterns around 60 EeV. Another possible interpretation of the observed shower development properties is a change in hadronic interactions above 100 TeV center of mass (TeV $\equiv 10^{12}$ eV), an order of magnitude higher energy than will be reached by the Large Hadron Collider (LHC) at CERN. A new puzzle is born: an injection at the source dominated by heavy nuclei is astrophysically unexpected, while significant changes in hadronic interactions represent novel particle physics.

To help discriminate between possible interpretations of recent results, we review in Section \ref{section:propag} the well-known physics of the propagation of ultrahigh energy cosmic rays: their interaction with the cosmic background radiation and the effect of cosmic magnetic fields. The effect of propagation on the observed spectrum, sky distribution, and composition depends on the source redshift evolution, the injected spectrum and composition, and the evolution of cosmic backgrounds and magnetic fields. The spectrum is cut-off due to photo-pion production of protons and photo-dissociation of nuclei off cosmic backgrounds. The composition simplifies to either proton or iron (or a mixture of the two) at trans-GZK energies. Anisotropies in the sky distribution of sources are blurred by magnetic fields for heavier primaries while protons keep most of the original anisotropies at trans-GZK energies.

Different scenarios for the transition between cosmic rays created in the Galaxy and those from extragalactic sources are discussed in Section \ref{section:transition}. Specific acceleration mechanisms envisioned for reaching these extremely high energies are the topic of Section \ref{section:accel}, including shock acceleration, unipolar inductors, and other proposals. In Section \ref{section:sources}, we survey known astrophysical sites that are reasonable candidates for UHECR sources, from compact objects such as, neutron stars (or magnetars), to gamma-ray bursts and active galaxies. Possible signatures of different candidate sources are discussed in light of future observations of UHECRs and other messengers of the extreme universe.

With a significant increase in the integrated exposure to cosmic rays above 60 EeV, next generation observatories may reach the sensitivity necessary to achieve charged particle astronomy and to observe ultrahigh energy photons and neutrinos, which will further illuminate the workings of the universe at the most extreme energies. We end with the ongoing and future search plans for the cosmic sources of ultrahigh energy particles.

Due to the limited space, we refer readers interested in details of observational techniques to \cite{Letessier11}, \cite{Beatty:2009zz}, \cite{Bluemer09}, and \cite{NW00}. Previous reviews on the astrophysics of UHECRs can be found in \cite{Cronin2005}, \cite{Olinto00}, \cite{B90}, and \cite{Hillas06,Hillas84}, while \cite{Bhatta00} also include a survey of UHECRs from cosmological relics. \cite{Stanev10} published a recent monograph on UHECRs while \cite{Gaisser91} covers cosmic rays of lower energies.  Recent reviews cover the closely related high energy neutrinos \citep{AM09} and high energy gamma-rays  \citep{Hinton09}.

\begin{figure}[!t]
\centerline{
\includegraphics[height=0.7\textwidth]{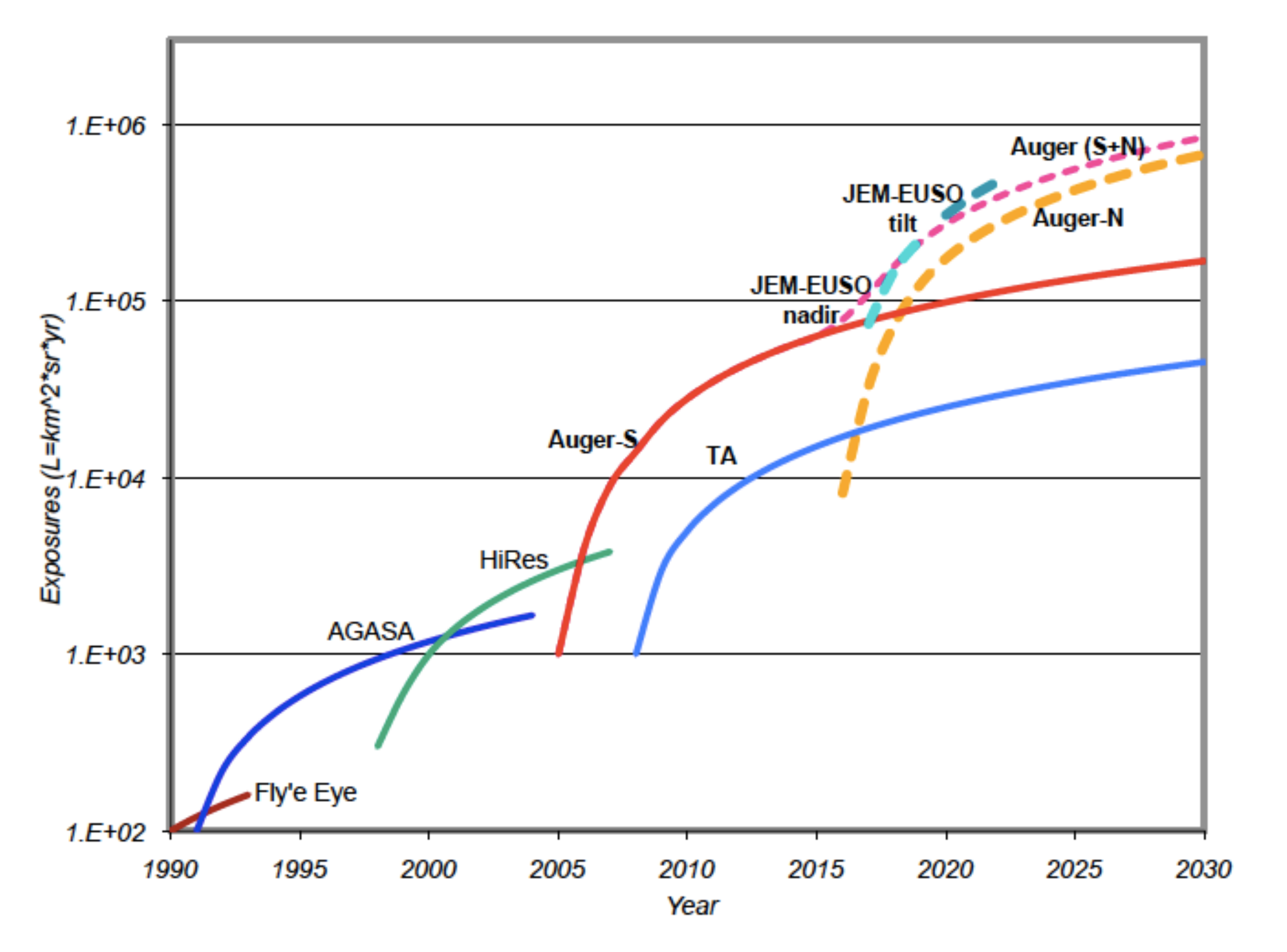}}
\caption{Evolution of the exposures of past, current, and planned UHECR observatories over time: Fly's Eye \citep{Baltrusaitis85}, AGASA \citep{Chiba92}, HiRes \citep{Boyer02}, Pierre Auger Observatory \citep{Abraham04}, TA \citep{TA1,TA2}. Projected exposures for Auger North \citep{AugerNorth} if construction start in 2016 and JEM-EUSO \citep{JemEUSO} if launched in 2017 including 20\% duty cycle.}
\label{fig:exposures}
\end{figure}

\section{Cosmic Ray Observations at Ultrahigh Energies}\label{section:observation}

After many decades of efforts to discover the origin of cosmic rays, current observatories are now reaching the necessary exposure to begin unveiling this longstanding mystery (see Figure~\ref{fig:exposures} for a the history of exposures for the largest observatories). The first detection of UHECRs dates back to \cite{Linsley63}, but it was only during the 1990s that an international effort began to address these questions with the necessary large-scale observatories. The largest detectors operating during the 1990s were the Akeno Giant Air Shower Array (AGASA), a 100 km$^2$ ground array of scintillators in Japan \citep{Chiba92}, and the High Resolution Fly's Eye (HiRes)  a pair of fluorescence telescopes that operated in Utah until 2006 \citep{Boyer02}. During their lifetimes, AGASA reached an exposure of $1.6 \times 10^3$ km$^2$ sr yr (or 1,600 $L$\footnote{In honor of UHECR pioneer John Linsley, we use the exposure unit  $L$ = 1 km$^2$ sr yr.}) while HiRes reached twice that. To date, the highest energy recorded event was a 320 EeV fluorescence detection \citep{Bird:1994} by the pioneer fluorescence experiment Fly's Eye \citep{Baltrusaitis85}.

Completed in 2008, the Pierre Auger Observatory is the largest observatory at present \citep{Abraham04}. Constructed in the province of Mendoza, Argentina, by a collaboration of 18 countries, it consists of a 3,000 km$^2$ array of water Cherenkov stations with 1.5 km spacing in a triangular grid overlooked by four fluorescence telescopes. The combination of the two techniques into a hybrid observatory maximizes the precision in the reconstruction of air showers, allowing for large statistics with good control of systematics. The largest observatory in the northern hemisphere, the Telescope Array (TA), is also hybrid \citep{TA1,TA2}. Situated in Utah, it covers 762 km$^2$ with scintillators spaced every 1.2 km overlooked by three fluorescence telescopes.

\subsection{Spectrum}\label{subsection:spectrum}

The observed cosmic ray spectrum (Figure \ref{figure1}) can be described by a broken power law, $E^{-s}$, with spectral  index $s = 2.7$ below the {\it knee} at $\sim$ 1 PeV ($=10^{15}$ eV) and $s\simeq 3$ between the knee and the ankle around 3 EeV \citep{KASCADE-Grande09}. Above the ankle, $s \simeq 2.6$  followed by the recently established flux suppression above about 30 EeV. With exposures around 10$^3 L$,  the measured spectra at energies where the GZK effect was anticipated had conflicting results: AGASA reported no flux supression at trans-GZK energies  \citep{Takeda98}, while early results from HiRes were consistent with the GZK prediction \citep{Abbasi04}. By 2006, HiRes accumulated enough statistics for the first significant observation of the GZK suppression  \citep{Abbasi08}, as displayed in Figure \ref{figure2}. This was confirmed by the Auger Observatory \citep{Abraham:2008ru} with a recent update starting at 1 EeV \citep{Abraham10} and based on $1.3\times10^4 L$ exposure (shown in Figure \ref{figure2}). The displayed error bars are statistical errors while the reported  systematic error on the absolute energy scale is about 22\%.  This systematic error allows for overall energy shifts that make the two observations consistent within the estimated errors. The highest energy event reported by Auger thus far is of 142 EeV \citep{Abreu10}.

\begin{figure}[!t]
\centerline{
\includegraphics[height=0.5\textheight]{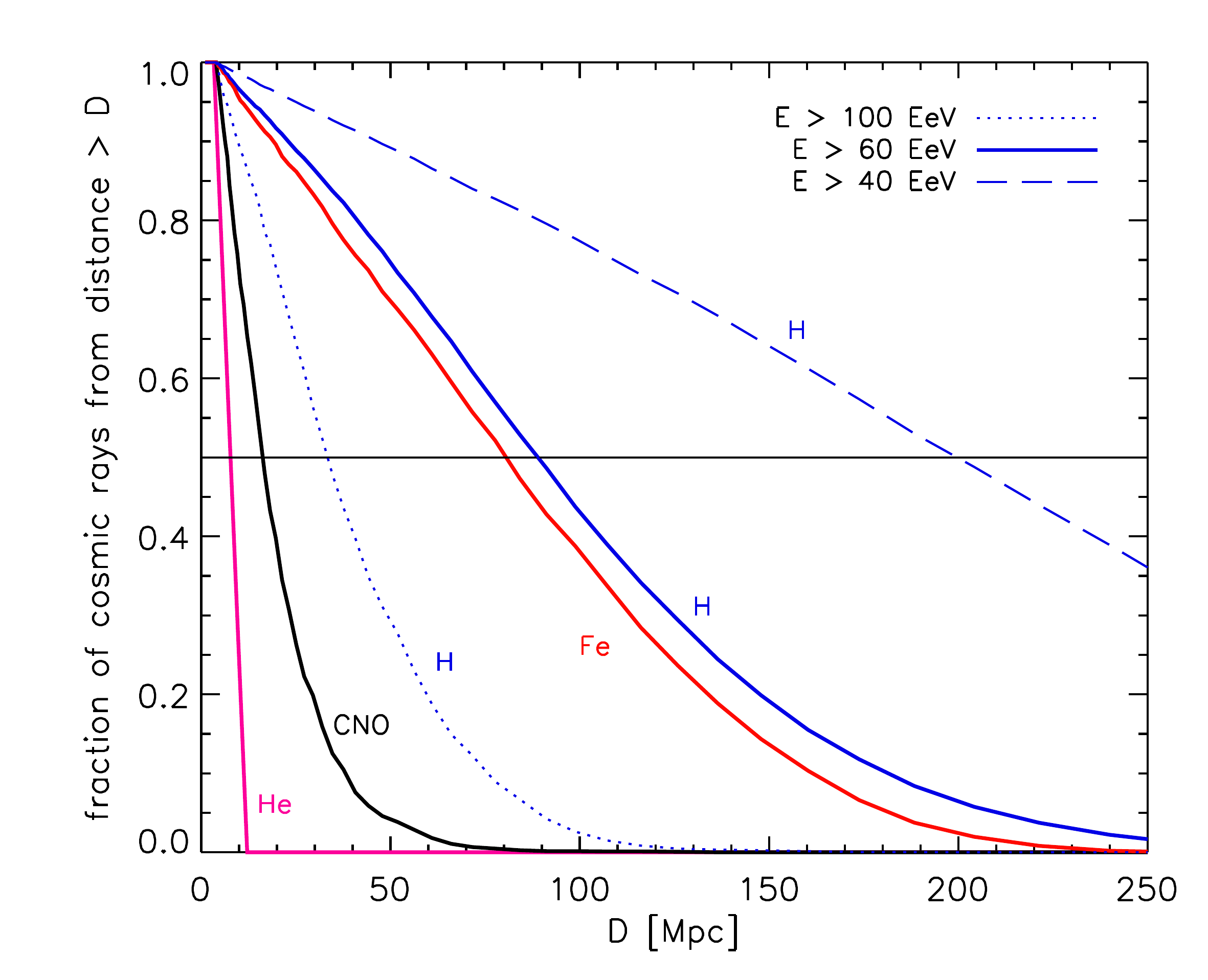}}
\caption{Fraction of cosmic rays that survives propagation over a distance $>D$, for protons above 40, 60, and 100 EeV and for He, CNO, and Fe above 60 EeV. Black solid line shows where 50\% of a given species can originate for a given atomic mass and energy. At trans-GZK energies ($E\gtrsim$ 60 EeV), only protons and iron survive the propagation over $D\gtrsim 50~$Mpc. Adapted from \cite{Allard07}.}
\label{figure3}
\end{figure}

Figure \ref{figure2} also shows the observed spectrum fit by different models of UHECR sources (taken from  \citealp{KAO10} and references therein). In the mixed composition and iron dominated models  \citep{Allard07}, the ankle indicates a transition from Galactic to extragalactic cosmic rays (see Section \ref{section:transition}), the source evolution is similar to the star formation rate (SFR),  and the injection spectra are relatively hard ($s\sim 2 -2.1$). In the proton dominated models in the figure, the ankle is due to pair production propagation losses \citep{BG88}, named ``dip models" \citep{BGG06}, and the injection spectra are softer for a wide range of evolution models. Models with proton primaries can also fit the spectrum  with harder injection with a transition from Galactic to extragalactic at the ankle \citep{WW04}.

The confirmed presence of a spectral feature similar to the predicted GZK cutoff, settles the question of whether acceleration in extragalactic  sources can explain the high-energy spectrum, ending the need for exotic alternatives designed to avoid the GZK feature. However, the possibility that the observed softening of the spectrum  is mainly due to the maximum energy of acceleration at the source, $E_{\rm max}$,  is not as easily dismissed. A confirmation that the observed softening {\it is} the GZK feature,  awaits supporting evidence from the spectral shape, anisotropies, and composition at trans-GZK energies and the observation of produced secondaries such as neutrinos and photons.

\subsection{Anisotropies in the Sky Distribution}

\begin{figure}[!t]
\centerline{\includegraphics[width=0.9\textwidth]{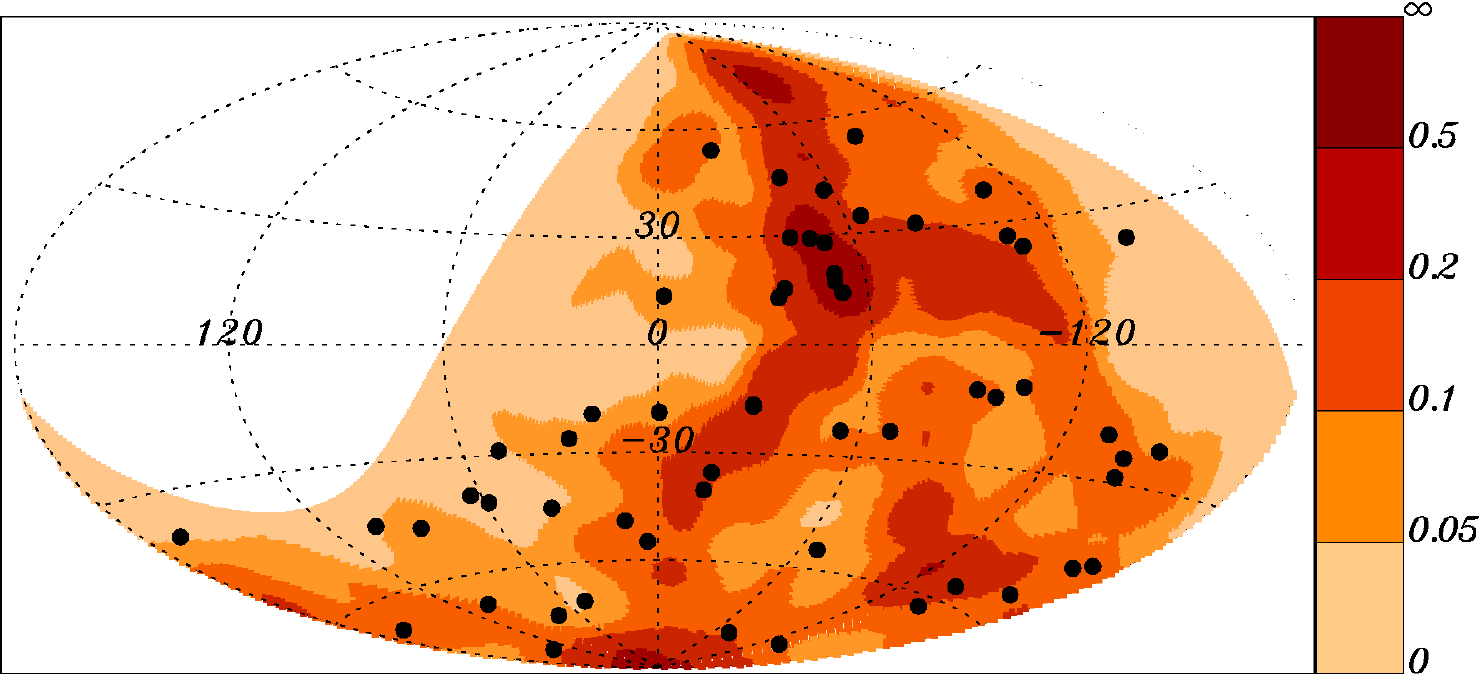}}
\caption{Arrival directions of cosmic rays with energy $E 
\ge 55$~EeV detected by Auger (black dots) in an Aitoff-Hammer projection of the sky in Galactic coordinates restricted to $|b|>10^\circ$  \citep{Abreu10}. 
Shaded areas represent a  smoothed  density map ($5^\circ$ smoothing angle) of the 2MRS galaxies within 200 Mpc over the Auger Observatory field of view.}
\label{figure4}
\end{figure}

The landmark measurement of a flux suppression at the highest energies encourages the search for sources in the nearby extragalactic universe using the arrival directions of trans-GZK cosmic rays. Above GZK energies, observable sources must lie within about 100 Mpc, the so-called GZK horizon or GZK sphere (\citealp{Harari06,Allard07}  and references therein). This effect is shown in Figure \ref{figure3} where the fraction of cosmic rays that arrive on Earth from a given distance is plotted for different energy protons ($>$ 40, 60, and 100 EeV) and for different nuclei (He, CNO, and Fe) arriving with energies above 60 EeV. At these trans-GZK energies, light composite nuclei are promptly dissociated by cosmic background photons (see Section \ref{section:propag}), while protons and iron nuclei  may reach us from sources at distances up to about 100 Mpc. Since matter is known to be distributed inhomogeneously within this distance scale, the cosmic ray arrival directions should exhibit an anisotropic distribution above the GZK energy threshold, provided intervening magnetic fields are not too strong. At the highest energies, the isotropic diffuse flux from sources far beyond this GZK horizon should be strongly suppressed.

Attempts to detect anisotropies at ultrahigh energies date back to the mid 1990s when hints of correlations with the local large scale structure  and with distant  BL Lacs objects were claimed and debated (see Section~\ref{subsection:astronomy}). With the increase in the number of observed ultrahigh energy events, these early claims have not been substantiated while different correlations have been recently reported.

The most recent discussion of anisotropies in the sky distribution of ultrahigh energy events began with the report that the arrival directions of the 27 cosmic rays observed by Auger with energies above 57 EeV exhibited a statistically significant correlation with the anisotropically distributed galaxies in the 12th VCV \citep{VC06} catalog of active galactic nuclei (AGN)  \citep{Auger1,Auger2}. The correlation was most significant for AGN with redshifts $z <$ 0.018 (distances $< $ 75 Mpc)  and within 3.1$^\circ$ separation angles. An independent dataset confirmed the anisotropy at a confidence level of over 99\% \citep{Auger1,Auger2}. The prescription established by the Auger collaboration tested the departure from isotropy given the VCV AGN coverage of the sky, not the hypothesis that the VCV AGN were the actual UHECR sources. In particular, a lack of events from the Virgo region showed that assuming the VCV AGN to be the sources gives a bad match to the observed event distribution \citep{Gorbunov08}.
No corresponding correlation was observed in the northern hemisphere by HiRes \citep{Abbasi08corr} where the distribution of their 13 trans-GZK events is consistent with isotropy.

Figure \ref{figure4} shows a map of the 69 events used in the recently published Auger update which includes another 42 trans-GZK events  \citep{Abreu10}.  With the new events, the correlation with the VCV catalog is not as strong for the same parameters as the original period (20 events correlate out of the original 27 while only 12 correlate out of the new 42). The data after the prescription period shows a departure from isotropy at the 3$\sigma$ level. With the currently estimated correlation fraction of 38\%, a 5$\sigma$ significance will require at least another four years of Auger observations \citep{Abreu10}.  

The VCV catalog is not a catalog produced by an instrument or survey strategy, but an extensive compilation of known AGN in the literature. A better set of catalogs which give a more homogeneous and statistically complete survey of the nearby universe over the large field of view of Auger has become recently available. In particular, the Swift-BAT catalog of AGN \citep{Swift1} and the 2MASS Redshift Survey (2MRS) catalog of galaxies \citep{2mass} are two catalogs where correlations may become more meaningful  \citep{Swift2,Abreu10}.  Figure \ref{figure4}  shows the example of Auger data superimposed on a density map generated with the 22,000 galaxies within 200 Mpc of the 2MRS catalog  \citep{Abreu10} with Galactic latitude $|b|>10^\circ$. The Auger trans-GZK events tend to align better with the distribution of  galaxies in 2MRS (and with Swift-BAT AGN) than with the isotropic scenario, however a significance of the anisotropy or a source class identification is hard to access with the current  limited statistics \citep{Abreu10}. 

Finally, the anisotropy reported by the test with the VCV catalog  may indicate the effect of the large scale structure in the distribution of source harboring galaxies or it may be due to a nearby source. An interesting possibility is the cluster of Auger events around the direction of Centaurus A, the closest AGN (at $\sim$  3.8 Mpc). The clustering around the Cen A region may also be due to the Centaurus cluster (as shown by the dark red region of Figure \ref{figure4}) which is much further away but on the general direction of Cen A \citep{KW08,KL08b}. Only much higher statistics will tell if Cen A is the first UHECR source to be identified.

\subsection{Composition}

\begin{figure}[!t]
\centerline{\includegraphics[height=0.7\textwidth]{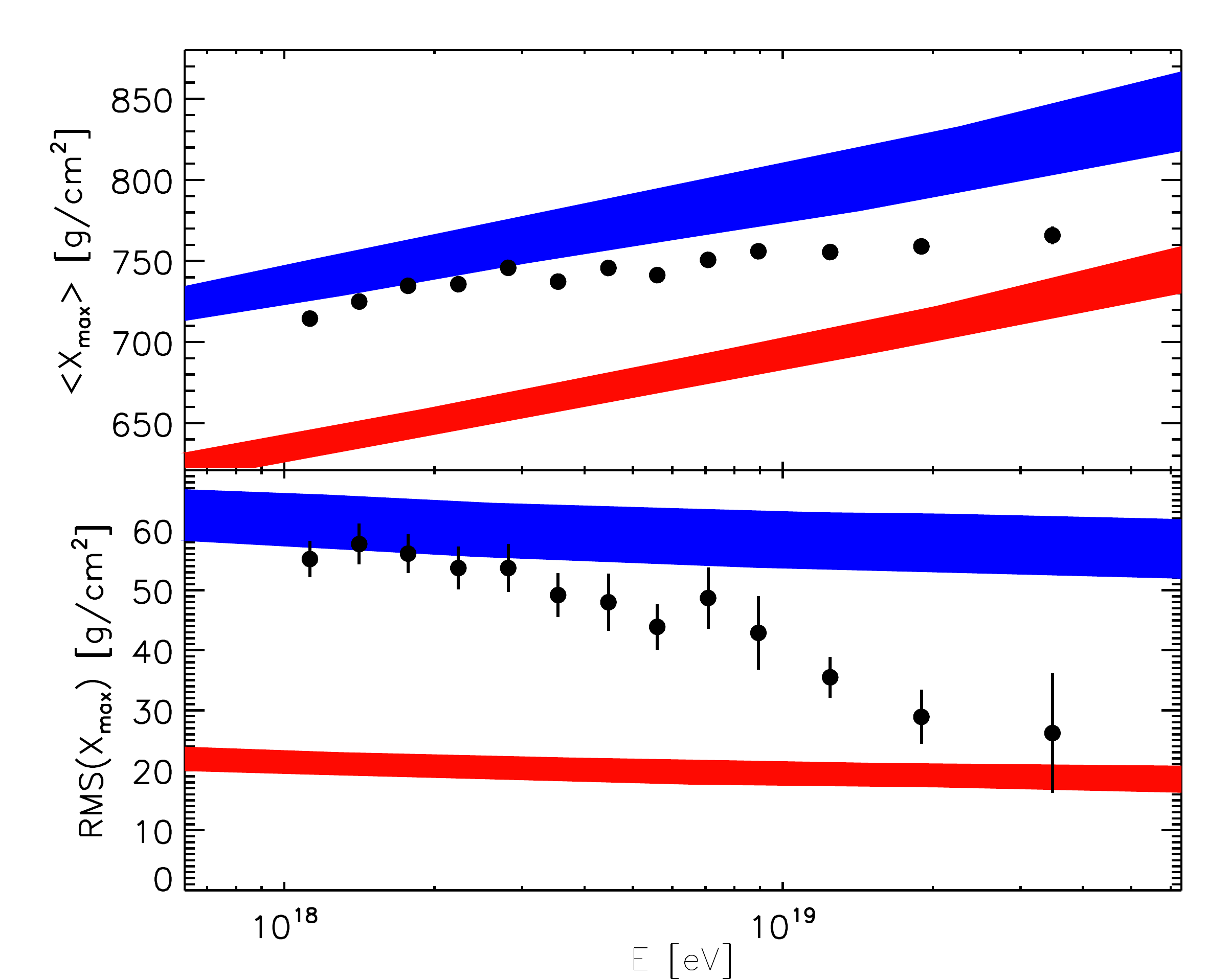}}
\caption{$\langle X_{\rm max}\rangle$ and RMS($X_{\rm max}$) as a function of primary energy, as measured by Auger fluorescence detectors \citep{Abraham:2010yv}. MC simulations from different hadronic interaction models are displayed for primary protons (blue) and primary iron nuclei (red).}
\label{figure5}
\end{figure}

The third key measurement that can help resolve the mystery behind the origin of UHECRs is their composition as a function of energy observed on Earth. Composition measurements can be made directly up to energies of $\sim$ 100 TeV with space-based experiments (see, e.g., \citealp{Ahn10}). For higher energies, composition is derived from the observed development and particle content of the extensive airshower created  by the primary cosmic ray when it interacts with the atmosphere. 

Presently, the best indicator of the composition of the primary particle is the depth in the atmosphere of the shower maximum, $X_{\rm max}$, given in ${\rm g/cm}^2$.  The average shower maximum, $\avg{X_{\rm max}}$, scales approximately as ${\rm ln} (E/A)$, where $E$ is the energy  and $A$ is the atomic mass  of the primary cosmic ray which generated the shower  (see, e.g., \citealp{Letessier11} and references therein). On average the shower maximum for protons occurs deeper in the atmosphere than that for the same energy iron nucleus, $\avg{X_{\rm max}^p} > \avg{X_{\rm max}^{Fe}}$. In addition, proton showers fluctuate more about $\avg{X_{\rm max}}$  providing another measure of composition, for example,  the root mean square fluctuations about $\avg{X_{\rm max}}$. Another useful measure of composition is the particle content of the shower such as the number of muons: proton showers have fewer muons than showers caused by heavier nuclei with the same energy. In practice, observed shower maxima and particle numbers are  compared with Monte Carlo airshower simulations which involve an extrapolation to higher energies of hadronic interactions known at energies of laboratory accelerators ($\lesssim$ TeV).

Observations of shower properties from the knee to just below the ankle indicate a general trend from light primaries dominating at the knee to heavier primaries dominating up to $\sim 0.1$ EeV (see, e.g., \citealp{Bluemer09}). These observations follow expectations that the knee is created by a rigidity\footnote{Rigidity is defined as particle momentum devided by charge, R $\equiv p/Z \propto E/Z$}  dependent end of Galactic cosmic rays which may be due to maximum acceleration at the sources and/or containment in the Galactic magnetic field. Just before the ankle, the trend seems to reverse back toward a lighter composition, being consistent with light primaries at 1 EeV as shown in Figure \ref{figure5}. Figure \ref{figure5} shows Auger data on $\avg{X_{\rm max}}$ and RMS($X_{\rm max}$) of 3754 events above 1 EeV together with a range covered by  simulations for protons and iron nuclei using different hadronic models \citep{Abraham10}. The dominance of light nuclei around a few EeV is also observed by HiRes who reported a final reconstruction of  815 events in \cite{Abbasi10}. 

A surprising trend occurs in Auger data above 10 EeV, a change toward heavy primaries is seen both in $\avg{X_{\rm max}}$ as well as in RMS($X_{\rm max}$) up to 40 EeV. As a mixture of different nuclei would increase the RMS($X_{\rm max}$), the observed narrow distribution argues for a change toward a composition dominated by heavy nuclei. Due to different reconstruction methods, the HiRes measurement of fluctuations is not easily displayed in Figure \ref{figure5}, but their data trend remains closer to light primaries up to around 50 EeV. The two results are consistent within quoted errors, so the situation is currently unclear. 

Shower properties observed by Auger up to 40 EeV are quite challenging to candidate acceleration models because they conflict with the prevailing view that the primaries are proton dominated. The observed hint of anisotropies also points to light primaries \citep{Abreu10}. Studies of ultrahigh energy nuclei propagation require unusual choices in attempts to fit  the observed composition indicators, such as a hard injection spectrum ($s \sim 1.6$) with primaries dominated by nitrogen or silicon  \citep{HT10}. Given the measurement uncertainties, some reasonable but ``disappointing'' options are also possible such as models where protons have a low maximum energy and the observed steepening of the spectrum, above 30 EeV, is due to the maximum energy of iron nuclei  \citep{Aloisio09,Allard08,Allard09}.  In this case, the feature in the observed spectrum is mainly due to the maximum accelerator energy which is coincidently close to the expected GZK cutoff. As discussed in Section \ref{section:sources}, most known astrophysical accelerators have $E_{\rm max}$ close to GZK energies, so the coincidence may actually be real.  The challenge to extragalactic candidate sources is to explain the origin of such high metallicities in the accelerated material. These observations have also motivated a return to the possibility that Galactic cosmic ray sources significantly contribute at ultrahigh energies  \citep{Calvez10}.

As recently highlighted in \cite{Schw10}, changes to hadronic interactions from current extrapolations provide a plausible alternative interpretation to the observed shower development behavior. Auger probes interactions above 100 TeV center of mass, while hadronic interactions are only known around a  TeV. The observation of anisotropies and secondary particles (neutrinos and gamma-rays) can lead to astrophysical constraints on the composition of UHECRs, opening the possibility for the study of hadronic interaction cross sections, multiplicities, and other interaction parameters at hundreds of TeV.

The detailed composition of UHECRs is still to be understood, but it is clear that primaries are not dominated by photons \citep{Aglietta:2007,Abraham:2009qb} or neutrinos \citep{Auger_nu09,Abbasi08neu}. Limits on the photon fraction place stringent limits on models where UHECRs are generated by the decay of super heavy dark matter and topological defects. Unfortunately, the uncertainties on the UHECR source composition, spectrum, and redshift evolution translates to many orders of magnitude uncertainty in the expected cosmogenic neutrino flux as discussed in Section \ref{section:sources}.

\section{The propagation of Ultrahigh Energy Cosmic Rays}\label{section:propag}

While propagating from their sources to the observer, UHECRs experience two types of processes: {\it (i)} interactions with cosmic backgrounds that affect their energy and their composition, but not their direction; and {\it (ii)} interactions with cosmic magnetic fields that affect their direction and travel time, but not their energy and composition. Both leave a variety of signatures on the  observables of UHECRs and generate secondary neutrinos and gamma rays (see Section \ref{section:multimess}).

\begin{figure}[!t]
\centerline{
\includegraphics[height=0.5\textheight]{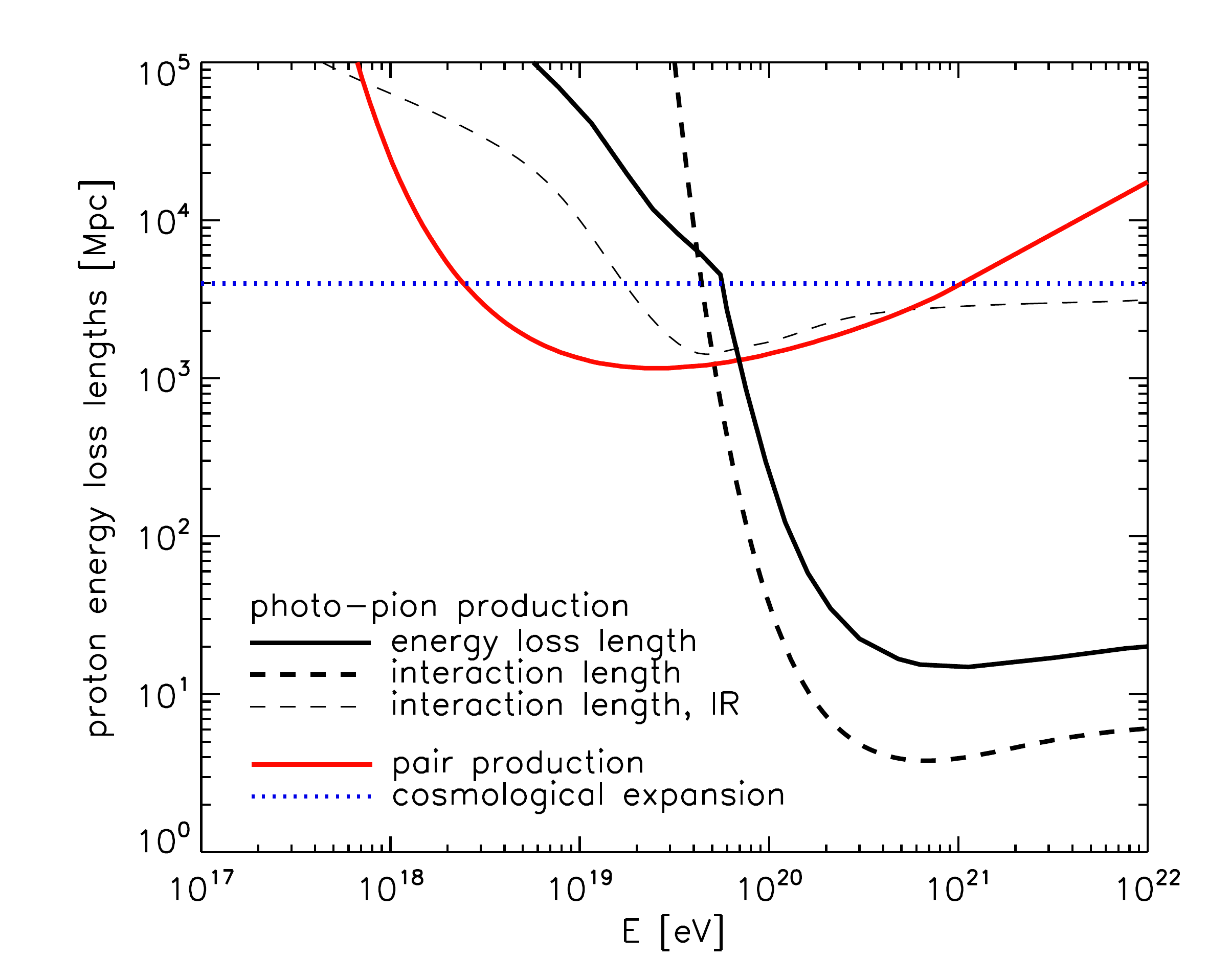}}
\caption{Proton energy loss lengths: black solid line for photo-pion production on CMB and IR-UV photons; red solid line for pair production on CMB photons. Dashed lines represent the interaction length (or mean free path to interaction) for photo-pion production on CMB photons (thick) and IR-UV photons (thin), assuming the background of \cite{SMS06}. The dotted line indicates the losses due to cosmological expansion.}
\label{figure6}
\end{figure}

\subsection{Interaction processes on cosmic backgrounds}
 
In the intergalactic medium, cosmic rays primarily interact with the Cosmic Microwave Background (CMB) photons at the highest energies, and with infrared,  optical, and ultraviolet background 
(IR-UV) photons at slightly lower energy (see, e.g., \citealp{Kneiske04,SMS06} for detailed background models).
 
Photohadronic interactions between protons and background photons mainly lead to pion production: $p\,\gamma \longrightarrow N+n \pi$ (here $N$ is a nucleon and $n$ is the number of pions produced), or to electron-positron pair production, also called Bethe-Heitler process: $p\,\gamma \longrightarrow p \,e^+\, e^-$. The energy threshold of these interactions for a photon of energy $\epsilon$ reads:
$E_{p,\rm \pi}  \sim 200\,\mbox{EeV}\, ({\epsilon_{\rm CMB}}/\epsilon)$
for pion production, and for pair production:
$E_{p,\rm ee} \sim 0.8\,\mbox{EeV}\,  ({\epsilon_{\rm CMB}}/\epsilon)$, with $\epsilon_{\rm CMB}\simeq 2.7 \,k_B T_{\rm CMB}\sim 6\times 10^{-4}$~eV, the mean energy of a CMB photon. 

We plot in Figure \ref{figure6} the energy loss lengths $x_{\rm loss}\equiv |E^{-1}{\rm d}E/c\,{\rm d}t|^{-1}$ for these two processes, using analytical formulae by \cite{Stecker68} (also see \citealp{Maximon68,Genzel73,Begelman90,Mucke99}). Above $E\sim 60$ EeV, the distance that particles can travel without losing their energy shortens considerably. 
 If cosmic rays originate from cosmological distances, their flux above this energy should be consequently suppressed, producing the well-known GZK feature in the UHECR spectrum (see Section~\ref{subsection:spectrum}). This property further imposes that the sources of the observed UHECRs at a given energy should be located in our local Universe, within a distance $l\lesssim x_{\rm loss}(E)$. Losses due to the cosmological expansion are also represented in Figure \ref{figure6}.

For primary cosmic rays with mass number $A>1$, different interaction processes come into play. 
At ultrahigh energies, nuclei photo-disintegrate on CMB and IR-UV photons through three main types of processes that contribute at increasing energy ranges: the Giant Dipolar Resonance (for $\epsilon \sim 8-30$~MeV), the Quasi Deuteron process (for $\epsilon \sim 20-150$~MeV), and the Baryonic Resonance (for $\epsilon \sim 150$~MeV). In a first approximation, one can consider that the Lorentz factor of the primary nucleus remains unchanged through these interactions. Nuclei also experience photo-pair production that decreases the Lorentz factor without affecting the number of nucleons. 

After pioneering work by \cite{PSB76}, energy losses for nuclei have been examined by several groups \citep{SS99,ER98b,ER98,Bertone02,Khan05,Allard05, Allard08, HTS05,HST08,ABG08}. One remarkable effect of the propagation of nuclei is that nuclei with mass number $A<20$ cannot travel farther than few tens of megaparsecs without disintegrating (see Figure \ref{figure3}). In particular, one can conclude that heavy nuclei could be found in abundance at trans-GZK energies only if the composition were essentially dominated by iron group nuclei. Such a composition can arise when the proton $E_{\rm max}$ is smaller than $E_{p,\pi}$, so that only heavy nuclei are present at greater energies \citep{Allard08,Aloisio09}. 

The effect of photo-hadronic interactions on the cosmic ray spectrum can be calculated analytically for protons \citep{BG88,BGG06}. Numerical codes such as SOPHIA \citep{Mucke99} enable the precise evaluation of the cross-sections for photo-hadronic interactions taking into account various channels, and of the produced flux of secondary particles (pioneered by \citealp{BG93}). Numerical Monte-Carlo methods are best suited to model inhomogeneous distribution of sources, calculate secondary emissions, and  treat the complex processes intervening in the propagation of nuclei in the intergalactic medium. Among the existing propagation codes that have been developed for this purpose, one might refer to the public code CRPropa \citep{Armengaud07} and to the complete nuclei propagation tool by \cite{Allard06}. 

The calculated spectra are in very good agreement with the observed spectra for a variety of chemical compositions, Galactic to extragalactic transition models, source evolution histories, and injection spectrum indices between $1.6-2.7$, for a fixed maximum acceleration energy, $E_{\rm max}$ (see, e.g., Figure \ref{figure2}). Kachelriess \& Semikoz (2006) demonstrate that relaxing the assumption of a single maximum acceleration energy and introducing a power-law distribution of $E_{\rm max}$ leads to a change in the overall propagated spectrum slope. A key region for models to fit is the ankle around a few EeV where the spectral slope changes (see Section \ref{section:transition}). The precise shape of the GZK feature depends on the local source density and on the transient or continuously emitting natures of the sources (see, e.g.,  \citealp{AB10,BGG06,Blasi99,MT98,MW96}). For instance, if $E_{\rm max}\gg$ 100 EeV a recovery of the spectrum at high energies can be observed by future detectors.

\subsection{The effects of Magnetic fields }

\begin{figure}[!t]
\centerline{\includegraphics[width=0.95\textwidth]{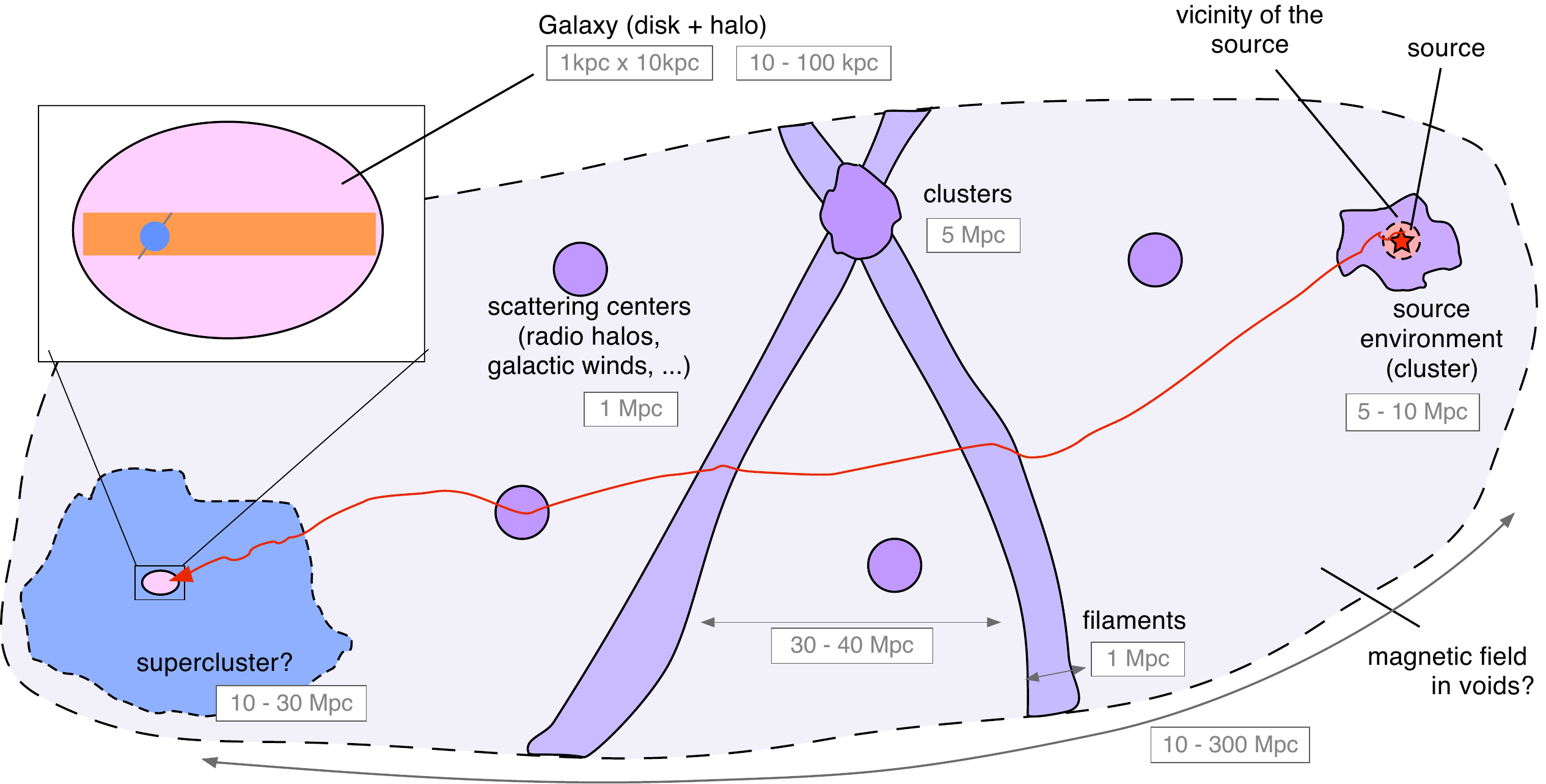}}
\caption{Schematic representation of magnetized regions intervening in UHECR propagation. Their approximative characteristic length scales are indicated in grey.}
\label{fig:mag_fields}
\end{figure}

The absence of powerful astrophysical counterparts in the arrival directions of UHECRs is probably related to the effect of cosmic magnetic fields that deflect and delay particles during their propagation. Charged particles are subject to the influence of magnetic fields in the source environment, in the intergalactic medium, and in the Galaxy, as depicted in Figure~\ref{fig:mag_fields}. Since very little is known about cosmic magnetic fields (for recent reviews see, e.g., \citealp{Beck08,KZ07,Vallee04,Widrow02,Kronberg94}), the parameter space for an accurate description is quite large.  Different propagation regimes  apply to different cosmic ray rigidities, from weak angular deflection at the highest energies or in weak magnetic fields, to the diffusive regime in strong fields or at low enough energies. Figure~\ref{fig:mag_fields} shows schematic representations of magnetized regions intervening in UHECR propagation. 

Much progress has been made in recent years on Galactic magnetic field observations (see \citealp{Han99,Han06,Han08,JFWE09}) and their effect on the propagation of UHECRs \citep{Harari99,AMES02,TT02,TT05}. These studies conclude that the deflection for particles of charge $Z$ and energy $E$ should not exceed $\sim10^\circ\, Z\,(40\,{\rm EeV}/E)$. In particular, the regular component of the Galactic magnetic field can distort the angular images of cosmic ray sources: the flux may appear dispersed around the source or globally translated in the sky with a small dispersion \citep{Harari99}. Source image distortions are stronger for heavier nuclei (see, e.g., \citealp{Giacinti10}). Since Galactic magnetic fields are not uniform in the sky, angular deflections also depend on the observed direction \citep{Harari99,KST07,TS08}. 

The extragalactic magnetic field (EGMF)\footnote{also called the intergalactic magnetic field (IGMF)} is much less known.  Measurements of Faraday rotation provide estimates of the magnetic fields in the core of clusters of galaxies, with typical strengths $\sim 1-40\,\mu$G (see above reviews). Outside clusters upper limits on the integrated strength of the magnetic field parallel to the line of sight, $B_{||}$, have been obtained using rotation measures \citep{Ryu98,Blasi99}: $\langle B_{||}^2\,\lambda_B\rangle^{1/2} \,\lesssim\,10^{-8}\,{\rm G\,Mpc^{1/2}}$, where the field reversal scale, $\lambda_B\lesssim 1~$Mpc, given the typical turbulent velocities (e.g., \citealp{WB99}). The future SKA project \citep{Beck07} will enlarge considerably the observations of the EGMF in our local Universe. Meanwhile, \cite{Neronov09} demonstrated that the observation of extended gamma-ray emission around point sources together with time delays in gamma-ray flares should provide robust measurements of the EGMF, and help constrain scenarios of its origin. 
Based on recent Fermi (Fermi Gamma-ray Space Telescope) observations, the existence of an EGMF of order $10^{?16}$ -- $10^{?15}$ G has been suggested using upper limits on the secondary emission of a few blazars \cite{Neronov10} and by the apparent (and debated, see e.g.,
 \citealp{Neronov11}) detection of pair halos in stacked images of a large number of sources \cite{Ando10}. 
These new developments may lead to stronger constraints or even detections of the EGMF, which at present can range from $10^{-16} - 10^{-9}$~G.

\begin{figure}[!t]
\centerline{\includegraphics[height=0.5\textheight]{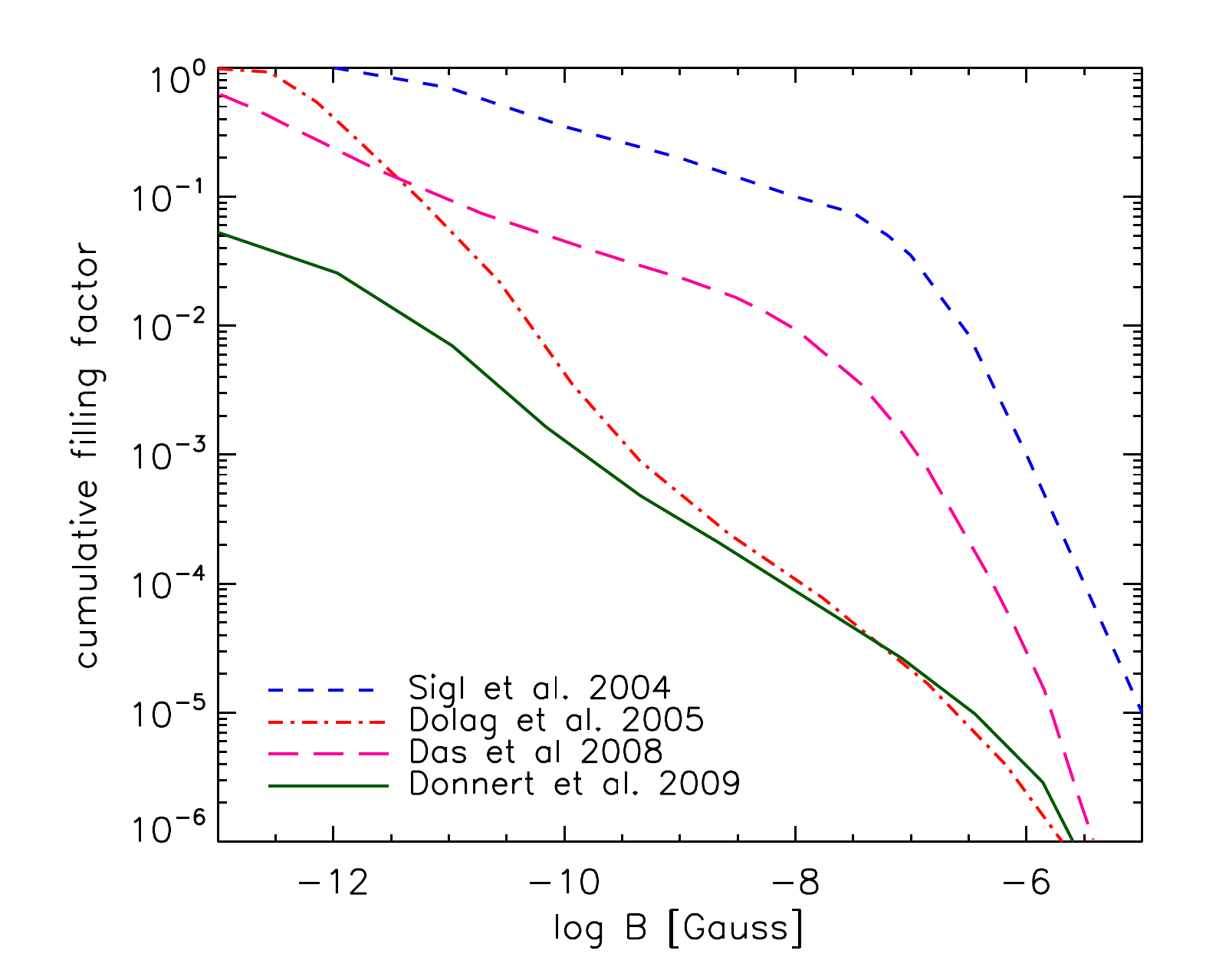}}
\caption{Cumulative volume filling factor of the extragalactic magnetic field for different numerical simulations. Blue dashed: \cite{SME04}, red dash-dotted: \cite{DGST05}, pink long dashed: \cite{Das08}, green solid: \cite{Donnert09}. }
\label{figure7}
\end{figure}

The origins of extragalactic magnetic fields are not well understood (e.g., \citealp{KZ07}). Some models set their origins in the primordial Universe (see \citealp{Widrow02} and references therein) other models  through magnetic pollution from astrophysical sources such as galactic winds or jets from radio-galaxies \citep{KLH99,Birk00,Aguirre01,Cen05,BSW05,BVE06,Scannapieco06}. If magnetic fields have a primordial origin, they should be all-pervading and amplified in dense regions by dynamical effects induced by large-scale structure formation. If galaxies are the primary generators of magnetic enrichment, the field should be rather concentrated in high peaked density regions and nearly suppressed in voids, depending on the efficiency of the winds \citep{BVE06}. A combination of the two scenarios may occur, as well as intermediate models in which magnetic pollution happens during the reionization epoch, generating slightly inhomogeneous fields already at high redshift. Whatever the origin of extragalactic fields, dynamical amplification in dense regions during structure collapse should play a key role in setting up their present configuration \citep{Bruni03,SG04,KC06,Ryu08}.  

The structure and strength of the EGMF can strongly affect the propagation of UHECRs. Early simulations of cosmic ray propagation in EGMFs considered homogeneous magnetic fields or  the lensing effect of the magnetized local supercluster \citep{Medina97,Lemoine97,SLO97,MT98,SLB99,Ide01,ILS02,IS02,Stanev03}. Analytical works of \cite{WW79,BO99,WM96,Achterberg99,Harari02a,Harari02b,KL08b}, and \cite{Aharonian10} have helped to establish the influence of the EGMF on observable UHECR quantities. 

More recently, various groups have developed simulations of the formation of large scale structures including magnetic fields, in order to model more realistic inhomogeneous configurations \citep{Ryu98,SME04,DGST04,DGST05,Bruggen05,Ryu08,DT08,Das08,Donnert09}. Most of these authors set an initial magnetic seed at high redshift which is then evolved in time. The overall amplitude of the field is rescaled at the end of the simulation so as to reproduce the observed strength of magnetic fields in the core of clusters of galaxies. \cite{Ryu98} and \cite{Das08} estimate directly the intensity of the magnetic field using the vorticity and the energy density calculated from the kinetic properties of the gas. With these methods, magnetic fields generated by astrophysical feedback can be hard to implement (see, however \citealp{Donnert09} who have incorporated some of these effects).

These simulations lead to very discrepant results concerning the configuration of the fields (see Figure \ref{figure7}). These differences probably stem from the different methods and assumptions made by each group, and illustrate the inherent complexity of these simulations. Indeed, \cite{Ryu98} and \cite{SME04} assume that magnetic seeds are generated by Biermann battery effects around accretion shocks, whereas \cite{DGST05} implement a homogeneous seed around redshift $z\sim 20$, and \cite{Donnert09} add to the latter method astrophysical magnetic pollution in localized spots at lower redshifts. The difference in configuration probably also results from the Eulerian and Lagrangian treatments used by the different groups and limitations on dynamical range and resolution.

Such discrepancies on the magnetic fields impact the range of deflections induced on UHECRs. For instance, \cite{SME04} find that protons with energy $E>100$~EeV should be deflected by $10-20^\circ$, while \cite{DGST04} find deflections of less than a degree at the same energy. 

\begin{figure}[tb]
\begin{center}
\centerline{\includegraphics[width=0.9\textwidth]{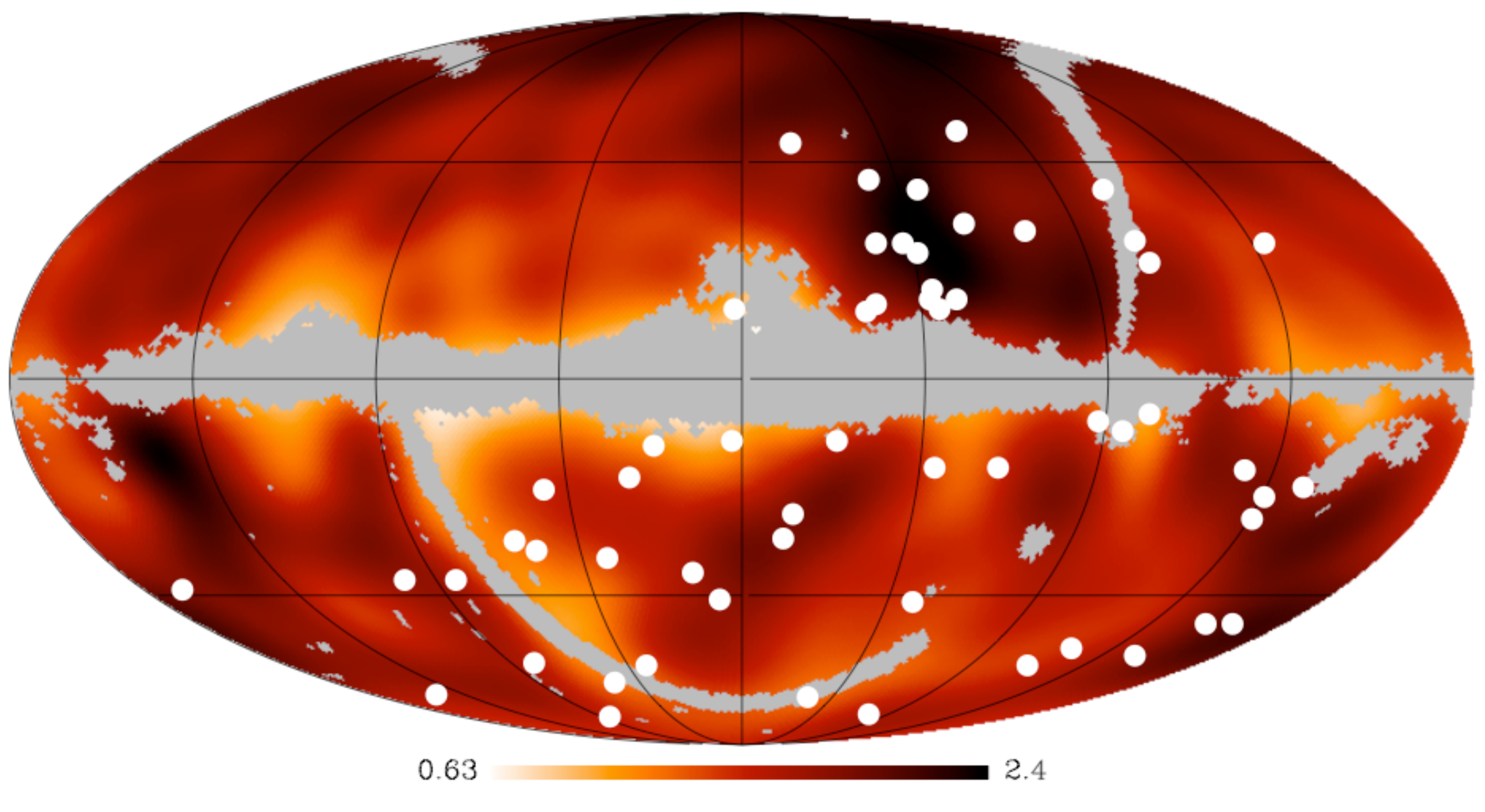}}
\caption{\footnotesize Expected angular deflection skymap (in degrees) for protons with energy $E\gtrsim 6\times10^{19}$~eV, calculated using the method developed in \cite{KL08b}. During their propagation, particles are assumed to be deflected on magnetized scattering centers that are distributed according to the galaxy density estimated using the PSCz catalog \citep{Saunders00}. The gray mask indicates the regions in the sky that are not covered by PSCz. Here, the parameters of the scattering centers (that sets a particular configuration of the EGMF) are chosen so as to reproduce the typical parameters found in the literature for magnetized filaments and radio halos: we assume that protons encounter $\sim 3$ scattering centers in average over a propagation distance of $\sim 100$~Mpc and experience a deflection of order $1.7^\circ$ after each encountering. White dots indicate the arrival direction of the cosmic rays of energy $E\gtrsim 6\times10^{19}$~eV detected by Auger \citep{Abreu10}.}
\label{fig:skymap}
\end{center}
\end{figure}

In the framework of UHECR propagation, a simpler and faster approach to model the inhomogeneous extragalactic magnetic field is to perform a scaling of the field strength to the underlying density field  \citep{MT97,TYS06,TS08,KL08a}. The assumed scaling law enables one to account for various types of magnetic field amplifications due to structure formation and for the high contrast fields that should be produced by astrophysical pollution. In addition, an analytic stochastic approach can be effective in describing  the propagation of UHECRs in the extragalactic fields because of the high energy of the particles and the low magnetization of voids  \citep{KL08b}.  Indeed, the deflection of UHECRs by magnetic fields of strength $B<10^{-12}$~G is lower than typical instrument resolutions which are $\sim 1^\circ$. Particle transport can then be viewed as a succession of rectilinear portions interrupted by deflections on localized magnetized regions (such as filaments, halos of radio-galaxies and galactic winds). This model can be applied to the coherent field amplified in numerical simulations as well as for the local enrichment processes due to astrophysical sources, and provides an effective framework to calculate the influence of magnetic fields on observable quantities of UHECRs  (see, for example, the expected angular deflection skymap for protons calculated for particular assumptions on EGMF presented in Figure~\ref{fig:skymap}).\\

If the magnetic field is sufficiently strong, or the particle energy ``not too high", UHECRs may enter a diffusive regime. One can define the scattering time of a particle, $t_{\rm scatt}$, as the timescale beyond which its quadratic deflection angle becomes of order unity  ($\delta \theta^2 \sim 1$). This quantity depends on the properties of the turbulence of the magnetic field. Diffusion occurs when the travel time exceeds $t_{\rm scatt}$, and the diffusion coefficients, which govern the transport on timescales $t\,\gg\,t_{\rm scatt}$, also depend on the properties of the turbulence. 

The propagated spectrum of UHECRs in the diffusive regime can be calculated analytically for a given injection spectrum, density and distribution of sources, for homogeneous magnetic fields \citep{AB04,L05,AB04}. Interestingly, \cite{AB04} have demonstrated that the analytical propagated spectra for both rectilinear (without magnetic field) and diffusive propagations converge when the typical distance between two sources is shorter than the characteristic propagation lengths (i.e. the scattering length and the energy loss lengths). This is particularly the case of continuously distributed sources for which the  diffusion of particles in the magnetic field does affect the final spectrum. 

When the strong magnetic field is not all pervading but limited to a particular geometry, the diffusive regime can lead to interesting confinement effects. For instance, the properties of UHECR transport in a strongly magnetized local supercluster or in clusters of galaxies have been discussed (e.g., \citealp{BGD90, BBP97,BO99,LSB99,Ide01,MT01}, and see Section~\ref{section:multimess} for more references on clusters of galaxies).

Another interesting aspect of diffusion, from the point of view of phenomenology, is that of magnetic horizons that will be discussed in the next section. If the length of the path travelled by the particle, from source to observer, becomes larger than the  Hubble length, $c H_0^{-1}$, the source cannot be seen in cosmic rays as it lies beyond the magnetic horizon. In a Hubble time, particles travel a linear distance $d\, \sim\, c\left(H_0^{-1}t_{\rm scatt}\right)^{1/2}\simeq 65\,{\rm Mpc}\,(ct_{\rm scatt}/1\,{\rm Mpc})^{1/2}$. Since $t_{\rm scatt}$, and hence $d$, increases with increasing energy, this produces a low-energy cut-off in the propagated spectrum.

\section{The Galactic to extragalactic transition}\label{section:transition}

The highest energy cosmic rays are likely to originate in extragalactic sources, given the strength  of Galactic magnetic fields and the lack of correlations with the Galactic plane. Low energy cosmic rays are easily created and contained in the Galaxy, so a transition region should occur in some intermediate energy.  ``A hypothesis blessed by long tradition is that'' Galactic cosmic rays end below 10 EeV ``and above that a different source is active (most plausibly in the nearby supercluster of galaxies)''  quoting \cite{Hillas84}. Modern measurements of the spectrum place a plausible transition region around the ankle at a few EeV (Figure \ref{figure1} and \ref{figure2}). However, the ankle can also be interpreted as the product of propagation losses due to pair production \citep{BG88,BGG06} in proton dominated scenarios allowing for a transition at lower energies.

The knee in the cosmic ray spectrum is likely to signal the $E_{\rm max}$ for light nuclei of dominant Galactic sources  and/or the maximum containment energy for light nuclei in the Galactic magnetic field. The same effect for heavier nuclei may cause  the softer spectrum above the knee (see, e.g., \citealp{L05,Hillas06}). Extragalactic sources producing spectra harder than $s=3$  can overtake the decaying Galactic  flux around the ankle.  Recent studies of a transition at the ankle which fit the observed spectrum and the composition trends in this energy region are discussed in \cite{Allard05} where different models are contrasted. Models based on proton primaries with a hard spectrum \citep{WW04}, on a mixed composition with proportions similar to the Galactic mix, or even on a composition dominated by heavy nuclei \citep{Allard07} fit well the UHECR spectrum and composition data around the ankle. In Figure \ref{figure2}, we show two examples of the so-called ``ankle transition models": one with source injection $s=2.1$, source composition similar to the Galactic mixture, and source evolution that follows the SFR; and a second model with similar source evolution and $s=2$, but a pure iron composition injected. Both models fit well the UHECR spectrum but predict different compositions throughout this energy range.

Ankle transition models work well for UHECR scenarios, but they were thought to challenge models for the origin of Galactic cosmic rays. The requirement that Galactic sources reach energies close to the ankle strained traditional models where  acceleration  in supernova remnants (SNRs) was expected to  fade around 1 PeV  \citep{LC83}. A modification to the traditional SNR scenario, such as magnetic field amplification in SN shocks \citep{BL01}, or a different progenitors  such as Wolf-Rayet star winds \citep{Bierm93}, and trans-relativistic supernovae  \citep{Budnik08} may explain the energy gap from PeV to EeV.  Taking into account magnetic field amplification and Alfvenic drift in shocks of Type IIb SNRs, \cite{Ptuskin10} find that Galactic cosmic ray iron can reach $E_{\rm max}\sim$ 5 EeV, allowing extragalactic cosmic rays to begin to dominate above the ankle.

The possibility that the ankle is due to pair-production losses during the propagation of extragalactic protons \citep{BG88}  has motivated an alternative model for the Galactic to extragalactic transition, called ``dip models"  \citep{BGG06}. The energy of the predicted dip is close to the observed ankle and a good fit to the spectrum over a large energy range is reached with a softer injection index as the Òdip protonÓ models shown in Figure \ref{figure2}. This option relaxes the need for Galactic cosmic rays to reach close to EeV energies, however it needs to be tuned to avoid strong (unobserved) spectral features between the knee and the ankle. Detailed models where the lower energy behavior of the extragalactic component blends smoothly with the Galactic cosmic rays have been developed using minimum energy and magnetic effects \citep{L05,AB05,Hillas06,KL08a,Globus08}.  
In some of these models a feature is produced around the ``second knee'' which may be observed around  0.5 EeV.  The dip model can fit the observed spectrum if the injection is proton dominated \citep{BGG05,Allard07} or with at most a primordial proton to helium mix \citep{Hillas06}, which gives a clear path for distinguishing it from mixed composition models. A proton dominated flux below the ankle region is a necessary condition for this model to be verified. 

Clarifying the structure of the transition region is crucial for reaching a coherent picture of the origin of Galactic and extragalactic cosmic rays.  This will require accurate spectrum and composition measurements from the knee to the ankle and beyond. KASCADE-Grande \citep{KASCADE-Grande09} has made great progress above the knee, while UHECR projects have started to lower their energy threshold such as the Auger Observatory enhancements \citep{AugerEnh09}: HEAT (High Elevation Auger Telescopes) and AMIGA (Auger Muons and InÞll for the Ground Array); and the Telescope Array Low Energy Extension (TALE) proposal. Having the same system covering a large range in energy will help control systematic offsets that degrade the accuracy of the needed precision. In addition, a strong multi-wavelength program  has shown that magnetic field amplification occurs in SNRs  and Galactic sources can reach further than previously believed. Finally, models of hadronic interactions will benefit from the energy reach of the LHC which can probe hadronic interactions at energies higher than the knee (Figure \ref{figure1}) and help constrain composition indicators between the knee and the ankle. 

\section{Acceleration Mechanisms}\label{section:accel}

The acceleration of charged particles is easily achieved in the presence of  electric fields. However, ubiquitous astrophysical plasmas destroy large scales electric fields throughout the universe.  Occasionally, high voltage drops that may lead to particle acceleration can be found in some regions of the magnetosphere or the wind of neutron stars, or near black holes and their accretion disks. In contrast, magnetic fields are omnipresent in astrophysical objects. Their variations in space and time imply the existence of transient electric fields which can supply a consequent amount of energy to charged particles. 

In our framework,  acceleration mechanisms must  fulfill the following criteria: they should enable charged particles to reach ultrahigh energies (from EeV to above 200 EeV) and the accelerated population should bear an injection spectrum (usually power-law) that would fit the observed UHECR spectrum after propagation. Below we summarize two of the most commonly cited acceleration mechanisms, Fermi acceleration and unipolar inductors, and briefly discuss other processes that have been proposed in the literature.

\subsection{Fermi acceleration at shock waves}\label{section:fermi}

The principle of Fermi acceleration is the transfer of energy from macroscopic motion to microscopic particles through their interaction with magnetic inhomogeneities. In the version elaborated by Fermi himself \citep{Fermi49}, magnetic scattering centers had random velocities which led to an energy gain of order $\Delta E/E\propto \beta^2$, where $\beta$  is  the average velocity of the scattering centers in units of $c$. This process is now called  2nd order Fermi acceleration, while the  1st order Fermi  process is the case where the macroscopic motion is coherent, such as a shock wave where particles can gain energy as they bounce back and forth, making  $\Delta E/E\propto \beta$ \citep{Axford77,Bell78,Blandford78}.
Shock waves are quite frequent in the Universe, for instance where an ejecta encounters the interstellar medium. SNR shocks are believed to be the sites where Galactic cosmic rays are accelerated via 1st order Fermi  processes. Popular shock regions for UHECR acceleration are GRB shocks, jets and hot spots of AGN, and gravitational accretion shocks. 

Second order Fermi  acceleration depends on the scattering time of particles in the magnetic turbulence and thus on the characteristics of the latter, which are poorly known in astrophysical objects. It is less efficient than 1st order Fermi in the non relativistic limit ($\beta\ll 1$). In the relativistic case \citep{Pelletier99}, it  has been applied to acceleration studies inside GRBs \citep{Pelletier00,Gialis03,Gialis04,Gialis05}.
 
First order Fermi  processes  under the `test particle' approximation\footnote{i.e., assuming that the density of accelerated particles is negligible compared to the thermal energy of the plasma in which the shock propagates, and hence that they do not produce a back-reaction on the shock.}, lead to simple power-law predictions for the spectrum of the accelerated population (see, e.g., \citealp{Gaisser91}). Several recent studies show that 1st order Fermi acceleration at relativistic shock waves is more intricate, even under the test particle approximation \citep{GA99,LP03,LPR06,Pelletier09}. \cite{LPR06} have pointed out that Fermi acceleration cannot happen if the Larmor radius of the accelerated particle is much smaller than the coherence length of the magnetic field. Indeed, the particle is then captured on a field line and advected far away downstream since the magnetic field is shock-compressed to a perpendicular configuration. Fermi acceleration should thus stop after a first back and forth cycle around the shock, unless the magnetic field is strongly amplified on spatial scales much smaller than the Larmor radius $r_{\rm L}$ of the particle. It was further demonstrated by \cite{Pelletier09} that the noise associated with the motion in the small scale turbulent magnetic field needs to overcome the unperturbed trajectory in the large scale coherent field, which sets an upper bound to $r_{\rm L}$ (see also, \citealp{Niemiec06}).

Such requirements being hardly reached for ultra-relativistic shocks, as discussed by \cite{Pelletier09}, one might advocate that the most efficient Fermi acceleration occurs around mildly relativistic shocks. The characteristics of the accelerated population strongly depend on shock parameters (e.g., the shock Lorentz factor $\Gamma_{\rm sh}$, temperature and pressure values). No general tendency is easily derived, as can be concluded from the large span of spectral properties obtained by the groups who have performed detailed studies of particle acceleration around mildly relativistic shocks in various situations (e.g., \citealp{BO98, Kirk00}). 

Accelerated particles can act as precursors of the shock and induce significant modifications of the gas flow upstream and downstream. The current of energetic particles can initiate plasma instabilities that tend to increase the level of magnetohydrodynamic (MHD) turbulence which scatter the particles. The possibility that streaming instabilities amplify the magnetic field, and that the acceleration efficiency could increase accordingly, was first discussed in the framework of non relativistic shocks in SNRs \citep{Bell78, LC83}. More detailed work was conducted by, e.g., \cite{McKenzie83,BE99,Lucek00,MD01,BL01,Bell04,Amato05,Amato06,AB09,Zirakashvili08,Reville08}. The estimated amplification of the upstream field is at most of two orders of magnitude. The generalization of the instability to the relativistic regime \citep{Milosavljevic06,Reville07,Pelletier09} reveals that this mechanism is not sufficient to allow efficient 1st order Fermi  acceleration around ultra-relativistic shocks. Other mechanisms outside the MHD range, such as Weibel like instabilities  \citep{Medvedev08} or resonant Cerenkov effects with plasma modes \citep{Pelletier09} might yet provide enough amplification to allow successful Fermi acceleration. Downstream of the shock, the relativistic two stream Weibel instability could amplify the magnetic field on small scales, to the level required by afterglow modeling of gamma-ray bursts \citep{Gruzinov99,Medvedev99}. In order to make progress on these issues, many groups are performing Particle-In-Cell simulations that endeavor to solve self-consistently the field-particle interactions (e.g., \citealp{Silva03,Hededal05, Dieckmann08,Spitkovsky08,Riquelme10}). It is promising that the latest simulations see evidences of Fermi acceleration and particle-wave interactions. 

Finally, when modeling Fermi acceleration around shocks, mechanisms of ``shear acceleration" have to be taken into account. 
Particles traversing a velocity gradient perpendicularly to a jet axis (instead of going along the axis as in the case of pure shock acceleration) can experience acceleration (e.g., \citealp{Rieger07}, see \citealp{Rieger05} for an application to acceleration of UHECRs in GRBs, \citealp{Lyutikov07} for a variant). This mechanism however depends on the characteristics of the plasma velocity gradients, which are not fixed, unlike the shock characteristics.

\subsection{Unipolar Inductors}\label{subsection:inductors}

Unipolar inductors have been suggested as alternative ways to accelerate particles to ultrahigh energies (see, e.g., \citealp{B90}). Below we focus specifically on neutron stars, but acceleration may also be caused by unipolar inductors in  other relativistic magnetic rotators, such as black holes with magnetized disks that lose rotational energy in jets. 

Rapidly rotating neutron stars generally create relativistic outflows (``winds"), where the combination of the rotational energy and the strong magnetic field induces an electric field ${\bf E} ={\bf v}\times {\bf B}/c$ (where {\bf v} and {\bf B} are the velocity and the magnetic field of the outflowing plasma). The wind thus presents voltage drops where charged particles can be accelerated to high energy. This model was first developed in the framework of ordinary pulsars (see, e.g., \citealp{Shapiro83} and references therein), but the latter do not supply enough energy to reach the highest energies ($E>10^{20}$~eV). \cite{Blasi00} pointed out that young neutron stars  with millisecond rotation periods $\Omega$ and very high surface magnetic fields $B_*$ (i.e., magnetars) could easily accelerate particles to: $E(\Omega)  \sim 3\times 10^{21} ~{\rm eV}\,Z \eta_1 ({B_*}/{2\times 10^{15}~{\rm G}})({R_*}/{10\, {\rm km}})^3({\Omega}/{10^4\, s^{-1}})^2$, where $\eta_1=0.1$ is the fraction of the voltage drop experienced by a particle and $R_*$ is the radius of the star. The spin down of the magnetar due to energy losses and the dependency of the particle acceleration energy on $\Omega$ lead to a power-law spectrum for the population of cosmic rays produced by a magnetar \citep{Blasi00,Arons03}:
${\rm d} N/{{\rm d} E} =  {9}{c^2I}( 1+{E}/{E_{\rm g}})^{-1}{(2ZeB_*R_*^3 E)^{-1}}$. $I$ is the principal inertial momentum of the magnetar and $E_{\rm g}$ is the energy corresponding to its angular velocity, at which gravity wave losses and electromagnetic losses are equal. \cite{Arons03} calculated that UHECRs should be produced at the very early stages of the lives of magnetars (after a few days to get down to $E\sim $ 6 EeV), hence their emission can be considered as an impulsive burst. Note that this model was introduced in the AGASA era, in order to explain the absence of the GZK cut-off; with our current data however, such a hard injection spectrum ($s=1$) is problematic as it does not fit the observed slope. An adequate distribution of initial voltages among magnetars can be found to soften the spectrum while leading to gravitational waves (Kotera 2011). 

The toy model described above needs to be further investigated on several issues, for instance on the nature of the ions injected in the wind, on the mechanism through which the current in the wind taps the available voltage, and on the escape of the accelerated particles from the wind and the wind nebula. A detailed discussion of such issues can be found in \cite{Arons03}.

\subsection{Other models}
Among other models that have been proposed in the literature, one might consider magnetic reconnection acceleration,  wake-field acceleration (often related to ponderomotive acceleration), and re-acceleration in sheared jets. In a plasma, a local reconfiguration of the magnetic field topology (or reconnection) happens when the plasma conductivity is not high enough to support the current associated with a magnetic field structure (see \citealp{Zweibel09} for a review). The field reaches a lower energy level configuration, and the liberated energy can be devoted to particle acceleration. 
This mechanism, responsible for the generation of high energy particles in solar flares, has been applied to UHECR acceleration in pulsar winds \citep{Coroniti90,Lyubarsky01}, in newborn millisecond pulsars \citep{Gouveia00}, in termination shocks of pulsar winds \citep{Lyubarsky03}, and in GRB outflows \citep{Thompson06}. Some authors also proposed that, in Poynting-flux dominated flows (e.g., flows with Poynting to kinetic flux ratio $\gtrsim 1$), 1st order Fermi  acceleration could aliment this mechanism, as particles reflected in the magnetized plasma could converge in the reconnection region \citep{Gouveia05,Giannios10}. 

A wake-field is created in a plasma when waves with high charge separation travel through the plasma. It leads to the formation of ponderomotive forces (longitudinal Lorentz invariant nonlinear forces that a charged particle experiences in an inhomogeneous oscillating electromagnetic field) that can accelerate particles if they are trapped in the wave: particles surf-ride the waves \citep{Tajima79,Chen02}. A simple form of surf-riding acceleration was developed in the case of a pulsar wind by \cite{Buckley77} and \cite{Contopoulos02}. A detailed discussion for magnetar winds can also be found in \cite{Arons03}.

Other acceleration mechanisms have been proposed and may contribute to the acceleration of cosmic rays in the Galaxy. These include a variety of second order processes and many of them can be observed to operate in solar physics. However, they are believed to be too slow to be relevant to the acceleration of UHECRs. 

\section{Candidate Sources  and their signatures}\label{section:sources}

The requirements for astrophysical objects to be sources of UHECRs are quite stringent. After reviewing some of the basic requirements, we briefly discuss plausible sources such as accretion shocks in large scale
structures, active galactic nuclei, gamma-ray bursts, and neutron stars or magnetars. For these different classes of candidate sources, we discuss the possibility of locating the sources with UHECR observations and review possible ways of
discovering the sources with secondary photons and neutrinos.

\begin{figure}[!t]
\begin{center}
\includegraphics[width=0.9\textwidth]{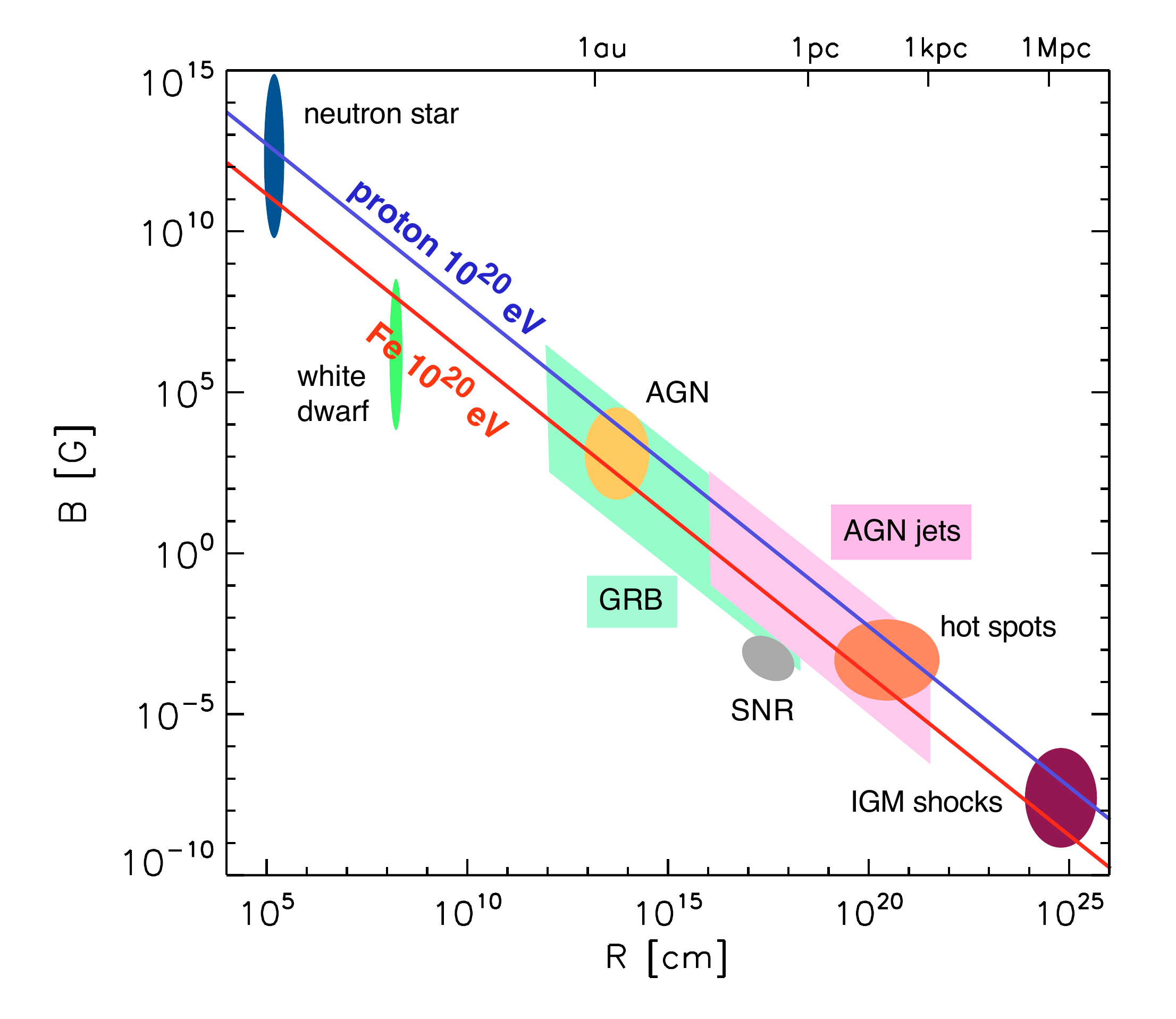}
\caption{Updated Hillas (1984) diagram. Above the blue (red) line protons (iron nuclei) can be confined to a maximum energy of $E_{\rm max}= 10^{20}$~eV.  The most powerful candidate sources are shown with the uncertainties in their parameters. }
\label{figure8}
\end{center}
\end{figure}

\subsection{Candidate Source Requirements}

The Larmor radius, $r_{\rm L} = E/ZeB \sim$ 110 kpc $Z^{-1} (\mu G/ B) (E/100$ EeV),  of UHECRs in Galactic magnetic fields is much larger than the thickness of the Galactic disk. Thus, confinement in the Galaxy is not maintained at the highest energies, motivating  the search for extragalactic sources.  Requiring that candidate sources be capable of confining particles up to $E_{\rm max}$, translates into a simple selection criterium for candidate sources with magnetic field strength $B$ and extension $R$  \citep{Hillas84}: 
$r_{\rm L} \le R$, i.e., $E \le E_{\rm max} \sim 1~\mbox{EeV}\ Z\,( {B}/{1~\mu\mbox{G}})( {R}/{1~{\rm kpc}} )$. 
Figure \ref{figure8} presents the so-called Hillas diagram where candidate sources are placed in a $B-R$ phase-space, taking into account the uncertainties on these parameters  (see also \citealp{Ptitsyna10} for an updated discussion on the Hillas diagram). Most astrophysical objects do not even reach the iron confinement line up to $10^{20}$~eV, leaving the best candidates for UHECR acceleration to be: neutron stars, Active Galactic Nuclei (AGN), Gamma Ray Bursts (GRBs), and accretion shocks in the intergalactic medium. The Hillas criterion is a necessary condition, but not sufficient. In particular, most UHECR acceleration models rely on time dependent environments and relativistic outflows where the Lorentz factor $\Gamma\gg 1$. In the rest frame of the magnetized plasma, particles can only be accelerated over a transverse distance $R/\Gamma$, which changes subsequently the Hillas criterion. 

The maximum accessible energy further depends on many details of the acceleration region but can be estimated by comparing the acceleration time, $t_{\rm acc}$, the escape time of particles from the acceleration region $t_{\rm esc}$, the lifetime of the source, $t_{\rm age}$, and the energy loss time due to expansion and to interactions with the ambient medium, $t_{\rm loss}$ (see, e.g., \citealp{Norman95,LW09}). The condition for successful acceleration can then be written $t_{\rm acc}\lesssim t_{\rm esc},t_{\rm age},t_{\rm loss}$. The escape timescale $t_{\rm esc} = R^2/(2D)$, where $D$ is the diffusion coefficient, depends on the characteristics of the transport of particles in the ambient medium, i.e., on the magnetic field and on the turbulence features. Detailed studies of this subject can be found in, e.g.,  \cite{Jokipii66,GJ99,CLP02, Yan02, CR04,Marcowith06}. Energy losses during acceleration are generally due to synchrotron radiation, to interactions with the radiative backgrounds, or to hadronic interactions, the latter process being mostly inefficient in diluted astrophysical media. The timescale for energy losses through synchrotron emission and pion production can be expressed in a generic way \citep{Biermann87}: $t_{\rm rad} = (6\pi m_p^4c^3/\sigma_{\rm T}m_e^2)E^{-1}B^{-2}(1+A)^{-1}$, where $A=240\,U_\gamma/U_B$ corresponds to the ratio of the energy density of radiation leading to pion production $U_\gamma$, to the magnetic energy density $U_B=B^2/8\pi$. In the central region of an AGN for example, assuming equipartition with the magnetic field (corresponding to the Eddington luminosity), for $E_{20}=E/10^{20}~$eV and $B_{\rm G}=B/1$G, $t_{\rm rad}\sim 10^5{\rm s}\,E_{20}^{-1}B_{\rm G}^{-2}$. This timescale has to be compared to the acceleration timescale which reads \citep{LW09}: $t_{\rm acc}={\cal A}\,t_{\rm L}$, where $t_{\rm L}$ is the Larmor timescale and ${\cal A}\gtrsim 1$ for all types of Fermi acceleration (non, mildly, or ultra-relativistic, 1st and 2nd order Fermi accelerations). For a non relativistic 1st order Fermi acceleration for instance, ${\cal A}\sim g/\beta_{\rm sh}^2$ and $t_{\rm acc}\sim 10^{7}{\rm s}\, g\,E_{20}B_{\rm G}^{-1}\beta_{\rm sh}^{-2}$, where the shock velocity $\beta_{\rm sh}\ll 1$ and $g\equiv D/(r_{\rm L}c)\gtrsim 1$. Majoring this timescale with the radiative loss timescale leads to a maximum acceleration energy in the central region of AGN of order: $E_{\rm max} \sim 10^{19}~{\rm eV}\, g^{-1/2}B_{\rm G}^{-1/2}\beta_{\rm sh}$.

In the generic case of acceleration in an outflow, \cite{LW09} compare this acceleration time and the dynamical time $t_{\rm dyn}\sim R/\beta_{\rm W}\Gamma_{\rm W}c$ of the outflow, to set a robust lower bound on the luminosity that a source must possess in order to be able to accelerate particles up to $E=10^{20}~{\rm eV}E_{20}$: $L>L_B\equiv \Gamma_{\rm W}R^2B^2/2>10^{45}\,Z^{-2}E_{20}^2~$erg~s$^{-1}$. The magnetic luminosity $L_B$ of the source is written as a function of the size of the acceleration region $R$ in the observer frame, in motion with Lorentz factor $\Gamma_{\rm W}$ (and velocity $\beta_{\rm W}$) and imparted with a magnetic field of characteristic strength $B$. This quantity is not straightforward to derive: the classical determination of the field strength using the synchrotron emission (assuming equipartition between the total energy density of non thermal particles and of the magnetic field for example), depends notably on the hardly known ratio between the leptonic and hadronic accelerated particles (e.g., \citealp{Beck05}). In the case of blazars for example, \cite{Celotti08} discuss that their jets are not magnetically dominated and that Faranoff-Riley I (FRI) radio galaxies, TeV blazars, and BL Lac objects only possess magnetic luminosities of order $10^{42-44}$~erg~s$^{-1}$.

It should be noted that the escape of particles from acceleration regions is an intricate issue that has been scarcely discussed in detail in the literature (note however the works of \citealp{Norman95,MPR01,Rachen08,AP09}). 
\cite{MPR01} and \cite{Rachen08} argue that one promising way to have high energy particles leave the magnetized acceleration site would be that they be transformed into neutrons. However, it is not obvious that such a scenario could produce a power-law spectrum; the shape of the spectrum depends on the evolution of the optical depth to photo-pion production (see, e.g., \citealp{W01}). 

In addition to being able to accelerate up to $E_{\rm max} >$ 200 EeV, candidate UHECR accelerators should have luminosities that can account for  the observed fluxes. A simple estimate of the required luminosity can be done 
assuming that all sources have the same injection spectral index $s$, a $L_{19}$ steady luminosity in cosmic rays above $E_{19}\equiv10^{19}$~eV, and that they are distributed homogeneously in the Universe with a density $n_{\rm s}$.  To account for the observed flux of UHECRs at $E_{19}$, the main quantity at play, $n_{\rm s}L_{19}$, scales as: $(E^3{\rm d}N/{\rm d}E)_{E=E_{19}} \sim 10^{24}~{\rm eV^2\,m^{-2}\,s^{-1}\,sr^{-1}}(n_{\rm s}/10^{-5}~{\rm Mpc^{-3}})(L_{19}/10^{42}~{\rm erg\,s^{-1}})$, for the case of $s=2.3$ and $E_{\rm max}=10^{20.5}$~eV. For reference, the number density of normal galaxies in the Universe today is of order $10^{-2}$~Mpc$^{-3}$ \citep{Blanton03}, and it drops to $10^{-9}-10^{-8}$~Mpc$^{-3}$ for Faranoff-Riley II(FRII) type galaxies \citep{Wall05}.
For transient sources, this scaling can be translated into: $(\dot{n}_{\rm s}/10^{-9}\,{\rm Mpc^{-3}\,yr^{-1}})$ $(E_{\rm tot,19}/3\times 10^{53}\,{\rm ergs})$, where $\dot{n}_{\rm s}$ is the birth rate of the source and $E_{\rm tot,19}$ the total injected energy in cosmic rays above $E_{19}$. 

Below we discuss the main astrophysical sites where UHECRs may originate.

\subsubsection{Gravitational accretion shocks}
The accretion of dark matter and gas produce shocks around the largest structures of the Universe (clusters of galaxies, filaments, walls), where diffusive shock acceleration can happen. For clusters of galaxies, one can estimate the linear extension of the magnetized shock to $\sim 1-10$~Mpc and the magnetic field intensity on both sides of the shock to $\sim 1~\mu$G (see, e.g., the recent radio detection of synchrotron radiation from bow shocks in a merging cluster by \citealp{vanWeeren10}, that indicates the presence of micro-gauss level magnetic fields far from the cluster center), which would allow particles to be confined up to $E\sim 10^{20}$~eV. Note however that the strength of the magnetic field {\it upstream} of the shock could actually be $\ll 1~\mu$G, as it was produced out of the weakly magnetized void; thus shock acceleration can occur only if the magnetic field upstream can be strongly amplified (M. Lemoine and E. Zweibel, private comm., and see tentative amplification mechanisms by \citealp{SS03,Zweibel10}). The detection of very high energy gamma rays from these shocks would enable us to constrain these parameters. Clusters of galaxies have been considered as promising accelerators by various authors: \cite{NMA95,KRJ96,KRB97,Miniati00,RK03,IAS05,Inoue07,MIN08}.   Recently, \cite{Vannoni09} performed a detailed time-dependent numerical calculation, including energy losses due to interactions of protons with radiative backgrounds and demonstrated that for realistic shock speeds of a few thousand km/s and a background magnetic field close to 1~$\mu$G, the maximum energy achievable by protons cannot exceed a few times $10^{19}$~eV, due to radiative losses.

\subsubsection{Active Galactic Nuclei}  AGN are composed of an accretion disk around a central super-massive black hole and are sometimes associated with jets terminating in lobes (or hot spots) which can be detected in radio. One can classify these objects into two categories: radio quiet AGN with no prominent radio emission or jets and radio loud objects presenting jets. Both categories could in principle accelerate particles in their nuclei: for a black hole of mass $M_{\rm bh}\sim 10^9~{\rm M}_\odot$, the equipartition magnetic field in the central region yields $B\sim 300$~G. Assuming the central region to be of order $R\sim 100$~A.U., particles could be confined up to $E_{\rm max} \sim $ 150 EeV and accelerated to $E\lesssim 10^{20}$~eV by electrostatic acceleration in the black hole magnetosphere (e.g., \citealp{Boldt99}). This energy is hardly reached by particles in practice due to energy losses that they experience in this dense region (see above and \citealp{Norman95,Henri99,Rieger00}). Radio loud galaxies could also accelerate particles in their inner jets, where one possible mechanism at play could be shear acceleration (see end of Section~\ref{section:fermi} and e.g. \citealp{Rieger07}). The quantity $B\ R\sim 0.3$~G~pc for the jets of $M_{\rm bh}\sim 10^9~{\rm M}_\odot$, leading to $E_{\rm max} \sim$ 300 EeV, but the acceleration is limited by photo-interactions and adiabatic losses making the escape of UHE particles non trivial \citep{Mannheim93}. The most powerful radio-galaxies (Faranoff-Riley II galaxies and their associated Flat Spectrum Radio Quasars, noted FSRQ) present hot spots and bow shocks, formed at the termination of the jets by interaction with the intergalactic medium. 
For these regions, the same estimates for particle confinement energies as in jets can be found. For hot spots, shock acceleration and escape should be easier than in the jet (see, e.g., \citealp{RB93}), but the acceleration in the bow shock is non trivial \citep{Berezhko08}. UHECR acceleration in AGN should lead to particular signatures in the gamma-ray spectrum of these sources, through various emissions such as proton synchrotron, photo-hadronic interactions that induce pair cascades, muon synchrotron etc. With more gamma-ray data on each source over a wide energy range (with, e.g., the future Cherenkov Telescope Array, or CTA), one could distinguish these hadronic signatures from leptonic acceleration ones and probe the UHECR acceleration in AGN \citep{Hinton09}. 
As discussed above, \cite{LW09} point out that only FRII/FSRQ radio-galaxies with magnetic luminosity $L_B\gtrsim10^{45}~$erg~s$^{-1}$ (following the modeling of \citealp{Celotti08}) meet the energetic requirements to accelerate particles to the highest energies.
Under this assumption, one may argue that the local FRII galaxies are well known and do not seem to correlate with the arrival direction of the highest energy events. Possible reasons for this could be that the extragalactic magnetic fields are stronger than expected, and/or that cosmic rays are heavy nuclei.
AGN are often suggested as continuous emitters of UHECRs, however transient events such as AGN flares more easily meet the UHECR acceleration requirements  \citep{Farrar09}. Other transient models for UHECR acceleration are discussed next.

\subsubsection{Gamma-ray Bursts} The explosion of a GRB leads to the formation of multiple shock regions which are potential acceleration zones for UHECRs. The value of the magnetic field at these shocks is estimated to be of order $B\sim 10^6~$G at a distance $R\sim 10^{12}$~cm from the center. These values are derived for internal shocks that happen before the ejected plasma reaches the interstellar medium, assuming $B\sim 10^{12}$~G near the central engine (of size $R\sim 10~$km) and an evolution $B\propto R^{-1}$. The wide green region presented in Figure \ref{figure8} stems from this dependency: parameters can take different values at different times of the GRB explosion. These objects have been examined by various authors as possible sources of UHECRs \citep{W95,W01,V95,Gialis03,Murase06,Murase08}. These authors invoke Fermi processes at external shocks \citep{V95}, at mildly relativistic internal or reverse shocks \citep{W95,W01,Murase08}, or 2nd order Fermi processes through multiple interactions with mildly relativistic internal shocks \citep{Gialis03}. Overall, models allow acceleration up to $\sim 10^{20}$~eV provided that some assumptions on the source parameters are verified (e.g., magnetic field strength, turbulence, etc.). The flux of gamma-rays reaching the Earth from GRBs is generally comparable to the observed flux of UHECRs, implying a tight energetic requirement  for GRBs to be the sources of UHECRs. With a GRB rate of $\sim 0.3$~Gpc$^{-3}$~yr$^{-1}$ at $z=0$, it can be calculated that the energy injected isotropically (regardless of beaming) in UHECRs is of order $E_{\rm UHECR}\gtrsim10^{53}$~erg, \citep{Guetta07,Zitouni08,Budnik08}. Note also that the transient nature of these objects could possibly explain the lack of powerful counterparts correlating with the arrival direction of the highest energy cosmic rays. 

\subsubsection{Neutron Stars} Neutrons stars -- young millisecond magnetars to be precise, easily fulfill the Hillas criterion, and might prove to be very good candidate sources, though they are scarcely discussed in the literature. 
Magnetars are neutron stars with extremely strong surface dipole fields of order $10^{15}$~G
(see Wood \& Thompson 2004, Harding \& Lai 2006 and Merghetti 2008 for reviews).  
\cite{Blasi00} studied the possibility that UHECRs are accelerated through unipolar induction in the relativistic winds for rapidly rotating magnetars, building up on previous constraints by \cite{Venkatesan97}.  The maximum energy reached by particles injected by these objects is very promising (see Section~\ref{subsection:inductors}).  \cite{Arons03} further developed the model and found that only 5\% of the extragalactic magnetar population need to be fast-rotators to account for the observed UHECR energetics. Magnetars, as GRBs, are transient sources and should not be observed in coincidence with UHECR arrival directions. The possibility of injecting large proportions of heavy nuclei into an acceleration region may be more easily met by young neutron stars than alternative models, due to their iron rich surface and early environment.

\subsection{Cosmic Ray Astronomy at Ultrahigh Energies}\label{subsection:astronomy}
One of the most puzzling facts concerning UHECRs is the absence of clear sources in the arrival directions of the highest energy events. Indeed, if  sources are powerful astrophysical objects, one would expect to see a counterpart in the arrival direction at the highest observed energies. At trans-GZK energies, current upper limits on the strength of cosmic magnetic fields suggest  that particles should not be deflected by more than a few degrees (unless they are heavy nuclei), thus some correlation should exist with the underlying baryonic matter. 

As a result, many authors have searched for correlations between existing data and astrophysical object catalogs. For example, \cite{TT01,TT02,Gorbunov02,Gorbunov05} found a correlation between HiRes event directions and BL Lac catalogs, that has been much debated \citep{EFS02,EFS04,TT04}. \cite{Stanev95} noted an association between arrival directions of UHECRs and the supergalactic plane that was not confirmed by AGASA results \citep{Takeda98} and that seems to have reappeared in the Auger results \citep{Stanev08}. The latest correlation result concerns the highest energy events ($E>$ 55 EeV) detected by the Auger Observatory and AGN within distance $<75$~Mpc \citep{Auger1,Auger2, Abreu10}. 

It must be underlined that the AGN correlating with the Auger events are mostly not very powerful Seyfert type galaxies, and are thus not favored as accelerators of UHECRs. Hence, what should be retained from that correlation is that it brings to light the hint of an anisotropic distribution of events at the highest energy with 99\% CL, its most reasonable interpretation then being that events trace the large scale structures along which AGN are distributed. Another possible  interpretation is that Auger may be observing in part the last scattering surface of UHECRs rather than their source population  \citep{KL08b}. The possibility that this `fake correlation' effect could play a non negligible role was shown numerically by \cite{Ryu10}.

Another explanation to the absence of counterparts in the arrival direction of UHECRs could also reside in the very nature of the sources. The delay induced by extragalactic magnetic fields of mean strength $B$ and coherence length $\lambda_B$ on particles of charge $Z$ and energy $E$ with respect to photons over a distance $D$ reads \citep{AH78}:
\begin{equation}\label{eq:delay}
\delta t\,\simeq\,2.3\times 10^2\,{\rm yrs}\,
Z^2\left(\frac{D}{10\,{\rm Mpc}}\right)^2\,\left(\frac{\lambda_B}{0.1\,{\rm Mpc}}\right)
\,\left(\frac{E}{10^{20}\,{\rm eV}}\right)^{-2}\left(\frac{B}{10^{-9}\,{\rm G}}\right)^2 .
\end{equation}
For intergalactic magnetic fields of lower overall strength ($B\lesssim 10^{-12}$~G), this formula indicates that the time delay be shorter than a year over 100~Mpc. However, the crossing of one single magnetized filament (size $\bar{r}_i$ and field strength $B$) will lead to a slight deflection that induces a time delay with respect to a straight line of order \citep{AH78,WM96,Harari02a}: $\delta t_i\,\simeq\, 0.93\times10^3\,{\rm yr}\,({\bar r_i/ 2\,{\rm
    Mpc}})^2 ({B_i/ 10^{-8}\,{\rm G}})^2\,({\lambda_i/  0.1\,{\rm Mpc}})({E/ 10^{20}\,{\rm    eV}})^{-2}$.

For transient sources like GRBs, neutron stars, or AGN flares which have an activity timescale $\ll \delta t$, this delay is sufficient to erase any temporal coincidence between UHECRs and their progenitors \citep{V95,W95}. Because these bursts are fairly rare in our local Universe (where observed UHECRs originate), one only expects to see particles from a particular bursting source if their arrival times are spread out over $\sigma_t \gtrsim 10^3$~yrs at $10^{20}$~eV. This spread should be particularly increased in the presence of magnetized scattering centers between the source and the observer. \cite{KL08b} and \cite{Kalli10} discuss that this effect could induce an artificially enhanced correlation of UHECR arrival directions with the foreground matter density, which could be measured to identify the transient nature of the sources. 

As discussed in Section 2.3,  UHECR sky anisotropies and their composition are tightly connected. In particular, if an anisotropy signal is measured above an energy $E_{\rm thr}$ assuming that it is produced by heavy nuclei of charge $Z$, one expects an anisotropy signal to be also present at energy $>E_{\rm thr}/Z$ due to the proton component, depending on the proton to heavy nuclei ratio $q_p/q_Z$ injected at the source \citep{LW09}. The evaluation of the anisotropy signals at both high and low energies should thus place constraints on the ratio $q_p/q_Z$.

\subsection{Multi-messenger approach}\label{section:multimess}

Secondary neutrinos and photons can be produced by UHECRs when they interact with ambient baryonic matter and radiation fields inside the source or during their propagation from source to Earth. These particles travel in geodesics unaffected by magnetic fields and bear valuable information of the birthplace of their progenitors. The quest for sources of UHECRs has thus long been associated with the detection  of neutrinos and gamma rays that might pinpoint the position of the accelerators in the sky.

The detection of these particles is not straightforward however: first, the propagation of gamma rays with energy exceeding several TeV is affected by their interaction with CMB and radio photons. These interactions lead to the production of high energy electron and positron pairs which in turn up-scatter CMB or radio photons by inverse Compton processes, initiating electromagnetic
cascades. As a consequence, one does not expect to observe gamma rays of energy above $\sim 100$~TeV from sources located beyond a horizon of a few Mpc \citep{WTW72,Protheroe86,PS93}. Above EeV energies, photons can again propagate over large distances, depending on the radio background, and can reach observable levels around tens of EeV \citep{Lee98}. Secondary neutrinos are very useful because, unlike cosmic-rays and photons, they are not absorbed by the cosmic backgrounds while propagating through the Universe. In particular, they give a unique access to observing sources at PeV energies. However, their small interaction cross-section makes it difficult to detect them on the Earth requiring the construction of km$^3$ or larger detectors (see, e.g., \citealp{AM09}).

Secondary neutrinos and gamma-rays  generated at UHECR sources have been investigated by a number of authors \citep{Szabo94,Rachen98,WB99,Mucke99,M00,Anchordoqui08,Kachelriess08,Ahlers09,AP09,MPR01} with particular focus on  emissions from AGN and  transient sources such as GRBs. The case of cluster accretion shocks has been studied by \cite{Inoue07} and \cite{MIN08}, and transient sources have been examined in details by \cite{WB00,Dai01,Dermer02,Murase06,Murase08}, and \cite{MI08} for GRBs and by \cite{Murase09} for magnetars. The normalization and the very existence of these secondaries highly depend on assumptions about the opacity of the acceleration region and on the shape of the injection spectrum as well as on the phenomenological modeling of the acceleration. For instance, \cite{WB99} obtain an estimate for the cosmic neutrino flux, by comparing the neutrino luminosity to the observed cosmic ray luminosity, in the specific case where the proton photo-meson optical depth equals unity. If the source is optically thick,  \cite{AP09} demonstrate that cosmic rays are not accelerated to the highest energies and neutrinos above $E\sim$  EeV are sharply suppressed. 

The existence of secondaries from interactions during the propagation of cosmic rays is less uncertain, but it is also subject to large variations according to the injected spectral index, chemical composition, maximum acceleration energy, and source evolution history. The magnetic field in the source environment, especially in clusters of galaxies, can play an important role by confining the charged UHECRs and thus leading to increased interaction probabilities \citep{BBP97, CB98, RGD04, demarco06, ASM06, MIN08, Wolfe08,KAM09}. 

\begin{figure}[!t]
\centerline{\includegraphics[height=0.6\textheight]{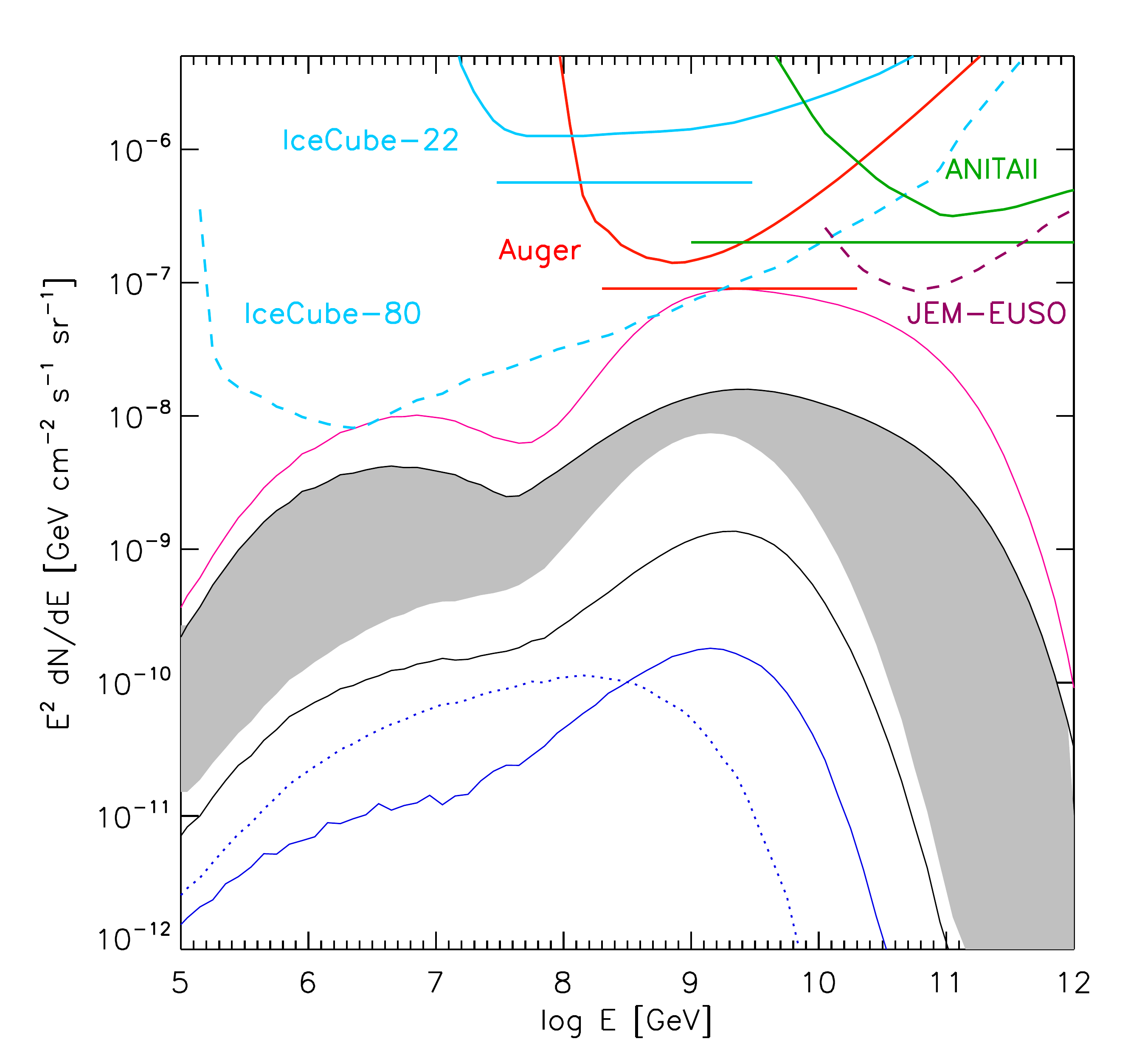}}
\caption{Cosmogenic neutrino flux for all flavors, for different UHECR parameters compared to instrument sensitivities. Pink solid line corresponds to a strong source evolution case (FRII evolution, see \citealp{Wall05}) with a pure proton composition, dip transition model, and $E_{\rm max}=$ 3~ZeV. Blue lines correspond to uniform source evolution with: iron rich (30\%) composition and $E_{Z,\rm max}<Z$ 10 EeV (dotted line) and pure iron injection and $E_{Z,\rm max}=Z$ 100 EeV (solid). Grey shaded range brackets dip and ankle transition models, with evolution of star formation history for $z<4$, pure proton and mixed `Galactic' compositions, and large proton $E_{\rm max} ( >  100$ EeV)). Including the uniform source evolution would broaden the shaded area down to the black solid line. Current experimental limits (solid lines) assume 90\% confidence level and full mixing neutrino oscillation. The differential limit and the integral flux limit on a pure $E^{-2}$ spectrum (straight line) are presented for IceCube 22 lines (pale blue, \citealp{Abbasi10}), ANITA-II (green, \citealp{ANITA10}) and Auger South (red,  \citealp{Auger_nu09}). For future instruments, we present the projected instrument sensitivities (dashed lines) for IceCube 80 lines (pale blue, acceptances from S. Yoshida, private communication, see also \citealp{Karle10}), and for JEM-EUSO (purple, \citealp{JemEUSO}).}
\label{figure9}
\end{figure}

A number of authors have estimated the cosmogenic neutrino flux with varying assumptions (e.g., \citealp{ESS01,Ave05,Seckel05,HTS05,Berezinsky06,Stanev06,Allard06,Takami09,KAO10}). Figure~\ref{figure9} summarizes the effects of different assumptions about the UHECR source evolution, the Galactic to extragalactic transition, the injected chemical composition, and $E_{\rm max}$, on the cosmogenic neutrino flux. It demonstrates that the parameter space is currently poorly constrained with uncertainties of several orders of magnitude in the predicted flux.
UHECR models with large proton $E_{\rm max} ( >  100$ EeV), source evolution corresponding to the star formation history or the GRB rate evolution,  dip or ankle transition models, and pure proton or mixed `Galactic' compositions are shaded in grey in Figure \ref{figure9} and give detectable fluxes in the EeV range with $0.06-0.2$ neutrino per year at IceCube and $0.03-0.06$ neutrino per year for the Auger Observatory. If EeV neutrinos are detected, PeV information can help select between competing models of cosmic ray composition at the highest energy and the Galactic to extragalactic transition at ankle energies.  With improved sensitivity, ZeV (=$10^{21}$ eV) neutrino observatories, such as ANITA and JEM-EUSO could explore the maximum acceleration energy. 

The detectability of photons from the electromagnetic cascade triggered by  pion production interactions has been addressed by several groups (e.g., \citealp{Lee98,Ferrigno04,ASM06,Gelmini07,KAL10}). The dilution of the cascaded signal -- due to the deflection of the electrons and positrons generated during the cascade -- depends on the assumptions made regarding the configuration of the EGMF. More specifically, the gamma-ray flux scales as the fraction of the line of sight in which the magnetic field is smaller than the value $B_\theta$ such that the deflection of the low energy cascade is $\theta$ \citep{KAL10}. For reference, $B_\theta\,\simeq\,2\times 10^{-14}\,$G for $\theta=1^\circ$. One can refer to Figure~\ref{figure7} for numerical estimates  of the magnetic  filling factor to arrive at the appropriate fraction for a specific line of sight and source location. Even under optimistic assumption on the magnetic field configuration, only sources with extremely high luminosities $L_{E,19}\gtrsim3\times10^{44}$~erg~s$^{-1}(d/100~{\rm Mpc})^{-2}$ and $L_{E,19}\gtrsim10^{43}$~erg~s$^{-1}(d/100~{\rm Mpc})^{-2}$ for  $E > 10^{19}\,$eV could be detected by current instruments such as H.E.S.S. and by the future CTA respectively, with fluxes of order $\sim 10^{-10}$~GeV~cm$^{-2}$~s$^{-1}$ around $1-10$~TeV \citep{Ferrigno04,KAL10}. 

\cite{GA05,GA07} argued that one should rather search for the GeV photons emitted by the synchrotron radiation of the secondary electrons, in presence of substantial magnetic fields in the source environment. Again, only the cases of rare powerful sources with cosmic ray luminosity $L_{E,19}>10^{44-46}$~erg$\,$s$^{-1}$ are promising in terms of detectability with both current and up-coming instruments. 
A source with a cosmic ray luminosity of $L_{E,19}\sim 10^{44}$~erg~s$^{-1}$ located at a distance $d\sim 100$~Mpc nearly overshoots the observed cosmic ray spectrum and is thus marginally excluded. Farther sources, with higher luminosity (e.g. $L_{E,19}=10^{46}$~erg~s$^{-1}$ at $d=1$~Gpc) would thus be more promising to observe in gamma-rays.  Such distant sources would contribute to about 10\% of the observed spectrum of UHECRs up to $E\sim 10^{19}$~eV, and the cosmic rays produced with higher energy would not reach the Earth due to energy losses.

Finally, sources located at a distance $\lesssim 10$~Mpc accelerating UHECRs should produce ultrahigh energy photons during their propagation, that can reach the Earth before experiencing Compton cascading. \cite{Taylor09}  studied this potential signature in the particular case of our closest radio-galaxy Cen~A (3.8~Mpc) and concluded that Auger should be able to detect $0.05-0.075$~photon per year from Cen~A, assuming that it is responsible for 10\% of the  cosmic ray flux above 60 EeV, and assuming a 25\% efficiency for photon discrimination.

One last messenger that is scarcely discussed in relation to UHECRs is gravitational waves. If anisotropy signals reveal that the source is of the transient type, one way to establish if UHECR sources are GRBs or magnetars would be to look for gravitational waves produced by the latter, as the former are believed to produce only faint signals below detectability (e.g., \citealp{Piran04}). \cite{K11} shows that the distribution of magnetar initial voltages required to reconcile the produced spectrum to the observed one, should lead to higher stochastic gravitational wave signals from these objects than previously calculated (e.g., \citealp{Regimbau08}). The observation of such a gravitational wave signal could be a probe that these objects meet the requirements (in terms of magnetization, rotation velocity, inertial momentum) to accelerate UHECRs to the highest energies.

It should be highlighted that due to the delay induced by EGMF on charged cosmic rays (see Section~\ref{subsection:astronomy}), secondary neutrinos, photons, and gravitational waves should not be detected in time coincidence with UHECRs if the sources are not continuously emitting particles, but are transient such as GRBs and young magnetars.

\section{The Search for Ultrahigh Energy Cosmic Ray Sources}\label{section:outlook}

The resolution of the long standing mystery of the origin of ultrahigh energy cosmic rays will require a coordinated approach on three complementary fronts: the direct ultrahigh energy cosmic ray frontier, the transition region between the knee and the ankle, and the multi-messenger interface with high-energy photons and neutrinos.

The most direct route to a resolution of this open question would be a precise measurement of the three pillars of UHECR observations: spectrum, anisotropies, and composition. The spectrum is much better measured today than just a few years ago and will certainly improve in the years to come with current observatories. The precise shape and energy scale of the ankle and the cutoff are excellent selectors of models. For instance, a possible recovery at the highest energies will clearly show the GZK nature of the cutoff as opposed to a cutoff caused by the maximum acceleration energy. A precise energy scale for the ankle will test a propagation dip versus a Galactic to extragalactic cosmic ray transition. 

More discriminating than the precise shape of the spectrum, but much harder to plan for success, is a clear observation of anisotropies. A nearby source, the first UHECR source, will clearly be a watershed in the field. A clear correlation with the large scale structure within 200 Mpc will also clear the path to zeroing in at the possible accelerators. 

The most difficult but key observable of the three pillars is a clear composition measurement. The dependence on hadronic models to translate shower properties into composition measurements make it difficult to reach clear conclusions. Great progress in this arena can be done at the transition region between the knee and the ankle, where the LHC has a direct access to the energy scale. Progress on enlarging the range of cosmic ray measurements to higher energies from the knee (Kascade-Grande) and to lower energies from the ankle (Auger HEAT and AMIGA, and TALE) will help construct a unified model of the cosmic ray properties in a region that hadronic models can be tested at the LHC\footnote{see http://www-ik.fzk.de/$\sim$needs/ for a list of measurements that can be made at accelerator programs to improve the ability of hadronic models to interpret cosmic ray observables.}.

A clear anisotropy determination, especially above 60 EeV, will help determine the composition astrophysically. At these energies possible composition mixtures simplify tremendously for sources above tens of Mpc into two options: protons or iron-like. Iron-like nuclei will have much broader spread in arrival directions due to known magnetic fields erasing small scale anisotropies and opening the possibility of a clear signature of proton primaries for some anisotropic patterns. An astrophysically determined primary composition will open a fruitful avenue to compare interaction models with observed shower properties at energies orders of magnitude above the reach of laboratory accelerators.

The spectrum and composition between the knee and the ankle should signal the transition from Galactic to extragalactic cosmic rays. A clear observation of a second knee or composition studies around the ankle can produce clear signatures of the transition. Anisotropies are not expected at these energies unless neutrons manage to escape acceleration sites. Progress in this energy range will also come from direct gamma-ray observations of cosmic accelerators (e.g., by Fermi, CTA, and HAWC) and their deeper understanding. Neutrinos in this energy range can also be crucial if sources can be identified or if cosmogenic neutrinos are observed in the PeV range (e.g., by IceCube or the future KM3NeT).

Observations of photons and neutrinos at ultrahigh energies will be extremely useful in distinguishing proposed scenarios (e.g., by IceCube, Auger, JEM-EUSO, and ANITA). Photons will only reach us from nearby accelerators, so they can be directly connected to acceleration sites. Neutrinos may be observed from nearby sources or from the diffuse background expected from the propagation of UHECRs from cosmologically distant objects. Having both a nearby view of the apparently brightest accelerators as well as the integrated flux from more distant objects will strongly constrain possible candidates. 

As an exercise, we end by contrasting two of the many possible outcomes of future observations. In outcome A, the large scale structure correlation is confirmed above the 5$\sigma$ level showing that UHECR sources exist within the GZK sphere and that their sky position is not smeared by more than a few degrees. These observations will imply that: 1) UHECR sources are more common than clusters of galaxies or powerful blazars, and 2) the composition above 60 EeV should be dominated by protons. If shower properties continue to show a trend toward iron-like behavior, it is likely to be due to changes in hadronic interactions, given that at 60 EeV only iron or protons arrive from sources further than tens of Mpc. This scenario is rich of multi-messengers: ultrahigh energy photons from nearby sources and neutrinos from sources and the diffuse background should be observable.  The type of galaxies that UHECR events correlate with may signal an origin in active galaxies, encouraging models based on central supermasive black holes. If starburst galaxies correlate better, magnetars and GRBs would be better candidate sources. Protons above 60 EeV do not prevent a mixed composition at EeV, so measurements of the composition at the ankle will help determine the injected composition and source spectral index.

In case B, above 60 EeV the spectrum does not show a recovery, anisotropies are not observed even after a significant increase in statistics, and shower properties indicate iron-like primaries. In this case, the maximum energy of the accelerators are likely to be  below GZK energies for protons and the spectrum cutoff is likely to be a combined effect of the maximum iron energy and the GZK effect. This coincidence is not an elegant solution but it is a clear possibility. A heavy composition at injection is more natural for models based on magnetars, while scenarios based on AGN and GRBs need to be modified to account for the suprisingly heavy composition. In this scenario, cosmogenic neutrinos and photons will not be easily detected leaving only the hope of observing a nearby source or of major technological advances to reach down orders of magnitude in flux.

Great progress in the ultrahigh energy cosmic ray frontier may lead to a completely different outcome than our speculative exercise above, but that will require a significant increase in statistics at trans-GZK energies. Current data suggest that watershed anisotropies will only become clear above 60 EeV and that very large statistics with good angular and energy resolution will be required. Auger will add $7\times10^3 L$ per year in the South while TA will add about $2\times10^3 L$ per year in the North. Current technologies can reach a goal of another order of magnitude if deployed by bold scientists over very large areas (e.g., Auger North). New technologies may ease the need for large number of detector units to cover similarly large areas. Future space observatories (e.g., JEM-EUSO, OWL, Super-EUSO) promise a new avenue to reach the necessary high statistics especially if improved photon detection technologies are achieved. With a coordinated effort, the next generation observatories can explore more of the $\sim 5$~million trans-GZK events the Earth's atmosphere receives per year.

\section{Acknowledgements}

We thank Denis Allard, Venya Berezinsky, Peter Biermann, Klaus Dolag, Ralph Engel, Susumu Inoue, Hyesung Kang, Martin Lemoine, Chris McKee, Teresa Montaruli, Sterl Phinney,  Dongsu Ryu, Dmitri Semikoz, Todor Stanev, and Ellen Zweibel for very helpful input to this manuscript. We thank the Auger collaboration for the brilliant work at the subject's frontier. This work was supported by  the NSF grant PHY-0758017 at  the University of Chicago, and the Kavli Institute for Cosmological Physics through grant NSF PHY-0551142 and an endowment from the Kavli Foundation.

\bibliography{UHECRreview}

\end{document}